\documentclass[fleqn,usenatbib]{mnras}

\usepackage{newtxtext,newtxmath}

\usepackage[T1]{fontenc}

\DeclareRobustCommand{\VAN}[3]{#2}
\let\VANthebibliography\thebibliography
\def\thebibliography{\DeclareRobustCommand{\VAN}[3]{##3}\VANthebibliography}

\usepackage{graphicx}	
\usepackage{amsmath}	
\usepackage{color}
\usepackage{multirow}
\usepackage{booktabs}
\usepackage{caption}
\usepackage{subcaption}
\usepackage{siunitx}
\usepackage{hyperref}

\usepackage[english]{babel} 

\usepackage{amsfonts,bm} 
\usepackage{soul} 

\DeclareSIUnit\sig{\sigma}

\def\Z#1{_{\lower1pt\hbox{$\scriptstyle#1$}}}
\def\Zu#1{^{\lower-1pt\hbox{$\scriptstyle#1$}}}
\def\ssp{\hspace{0.09em}}
\def\sp{\hspace{0.06em}}
\def\mbf#1{\mathbf{#1}}
\def\mrm#1{\mathrm{#1}}
\def\rme{\mathrm{e}}
\def\im{\mathrm{i}}
\def\Pr{p}
\def\be{\begin{equation}}\def\ee{\end{equation}}
\def\bea#1\eea{\begin{align}#1\end{align}}
\def\mid{\,|\,}

\def\LCDM{$\Lambda$CDM}
\def\OmO{\Omega_{m0}}

\def\qso{{\rm{Q}}}
\def\bqso{b_\qso}
\def\dqso{\delta_\qso}
\def\T{\intercal}
\def\healpix{{\sc HEALPix}}
\def\npix{{n_{\rm pix}}}
\def\dif{\mathrm{d}}
\def\mat#1{{\mathbf{#1}}}
\def\x{\mathbf{x}}

\def\v{\mathbf{v}}

\def\bk{\mathbf{k}}
\def\n{\hat{\mathbf{n}}}
\def\cov{\mathbf{C}}

\def\calL{\mathcal{L}}
\def\calN{\mathcal{N}}

\def\kin{{\rm k}}
\def\clu{{\rm c}}
\def\ckin{A_\kin}
\def\dkin{\mathbf{d}_{\kin}}
\def\dint{\mathbf{d}_{\clu}}

\def\Dkin{D_{\kin}}
\def\Dint{D_{\clu}}
\def\Dtot{D}
\def\sigkin{\sigma\Z{\kin\kin}}
\def\sigint{\sigma\Z{\clu\clu}}
\def\sigkinint{\sigma\Z{\kin\clu}}

\newcommand{\dotdeg}{\rlap{.}^\circ}
\def\PR{\mathcal{P}}

\title[Bayesian Analysis of the Number-Count Dipole]{Testing the Cosmological Principle with CatWISE Quasars: \\ A Bayesian Analysis of the Number-Count Dipole}

\author[Dam, Lewis \& Brewer]{
Lawrence Dam$^{1,2}$,\thanks{E-mail: \href{mailto:lawrence.dam@unige.ch}{lawrence.dam@unige.ch}} 
Geraint~F.~Lewis$^{1}$\thanks{E-mail: \href{mailto:geraint.lewis@sydney.edu.au}{geraint.lewis@sydney.edu.au} (GFL)}
\&
Brendon~J.~Brewer$^{3}$
\\
$^{1}$Sydney Institute for Astronomy, School of Physics, A28, The University of Sydney, NSW 2006, Australia\\
$^{2}$D\'{e}partement de Physique Th\'{e}orique and Center for Astroparticle Physics, Universit\'{e} de Gen\`{e}ve, 24 quai Ernest-Ansermet, 1211 Gen\`{e}ve 4, Switzerland\\
$^{3}$Department of Statistics, The University of Auckland, Private Bag 92019, Auckland 1142, New Zealand\\
}

\date{Accepted XXX. Received YYY; in original form ZZZ}

\pubyear{2023}

\begin{document}
\label{firstpage}
\pagerange{\pageref{firstpage}--\pageref{lastpage}}
\maketitle

\begin{abstract}
The Cosmological Principle, that the Universe is homogeneous and isotropic on sufficiently large scales, underpins the standard model of cosmology. 
However, a recent analysis of 1.36 million infrared-selected quasars has identified a significant tension in the amplitude of the 
number-count dipole compared to that derived from the CMB, 
thus challenging the Cosmological Principle.
Here we present a Bayesian analysis of the same quasar sample, testing various hypotheses using the Bayesian evidence. 
We find unambiguous evidence for the presence of a dipole in the distribution of quasars with a direction that is consistent with
the dipole identified in the CMB. However, the amplitude of the dipole is found to be 2.7 times
larger than that expected from the conventional kinematic explanation of the CMB dipole,
with a statistical significance of $5.7\sigma$.
To compare these results with theoretical expectations, we sharpen the \LCDM\ predictions for the probability distribution of the
amplitude, taking into account a number of observational and theoretical systematics. In particular, we show that the presence of the Galactic plane mask causes a considerable loss of dipole signal due to a leakage of power into higher multipoles, exacerbating the discrepancy in the amplitude. By contrast, we show using probabilistic
arguments that the source evolution of quasars improves the discrepancy, but only mildly so.
These results support the original findings of an anomalously large quasar dipole, independent of
the statistical methodology used.
\end{abstract}




\section{Introduction}
The Cosmological Principle, the idea that the Universe is spatially homogeneous
and isotropic when viewed at sufficiently large scales, underlies the use of
Friedmann--Lema\^{i}tre--Robertson--Walker (FLRW) world models in the standard concordance cosmology, \LCDM.
In these models there exist ideal observers for whom their view is an isotropic universe, such that
in this `cosmic rest frame' the CMB appears maximally isotropic~\citep{2011RSPTA.369.5115M}.
Any observer moving with velocity $\v$ relative to this frame
will observe a dipole anisotropy in the CMB temperature, $\Delta T/T\simeq\bm\beta\cdot\n$, where $\bm\beta=\v/c$
and $\n$ is the direction of observation~\citep{1967Natur.216..748S,1968PhRv..174.2168P}.
The fact that the  CMB dipole as observed from in the heliocentric frame (which is about a
hundred times larger than the primary anisotropies) 
is conventionally taken as evidence that the solar system 
is moving with speed~\citep{2020A&A...641A...3P}
\be\label{eqn:vel-hel}
    v=(369.825\pm0.070)\,{\rm km\,s}^{-1}
\ee
towards
\be
    (l,b)=(264\dotdeg021\pm0\dotdeg009,\,48\dotdeg253\pm0\dotdeg004),
\ee
relative to the CMB rest frame.

To test whether the CMB dipole has a genuine kinematic origin, \citet{1984MNRAS.206..377E}
proposed a simple consistency test using the
number counts of radio sources on the sky:
Given an isotropic distribution of sources, forming a  background of uniform emission,
our putative velocity should induce in the number counts a dipole anisotropy
of the amplitude and direction expected by equation~\eqref{eqn:vel-hel},
{\it if the kinematic interpretation is correct}.
Supposing a population of radio sources with identical flux density spectra
$S\propto \nu^{-\alpha}$ (where $\nu$ is the frequency and $\alpha$ the spectral index),
and integral source count above flux density threshold $S_*$ given by
$\dif N(>S_*)/\dif\Omega\propto S_*^{-x}$, \citet{1984MNRAS.206..377E} showed that
the number counts across the sky exhibits a dipole anisotropy $\Delta N/N=\dkin\cdot\n$ with
\be
    \dkin=[2+x(1+\alpha)]\bm\beta.
    \label{eqn:d-kin}
\ee
This is the \emph{kinematic dipole}.
Here ${\bm\beta} = \v/c $, where $\v$ is the velocity of the heliocentric frame relative to the
`matter rest frame', the frame in which the radio sources are observed at rest.
Unless we have reason to expect, as in the standard model, that the matter rest frame
coincides with the CMB rest frame, there is no {\it a priori} reason that the velocity
in equation~\eqref{eqn:d-kin} is the same as the velocity in equation~\eqref{eqn:vel-hel}.
Conversely, the agreement of these two velocities
then provides an indirect check on the Cosmological Principle
\citep[for a recent appraisal on its observational status, see][]{Aluri:2022hzs}.

In performing this check, one important advantage
is that $\dkin$ is independent of cosmological
assumptions; in principle, $\dkin$ arises from special-relativistic considerations, namely, the aberration and
Doppler effects. 
In particular, for $\beta=370\,{\rm km\,s}^{-1}/c\approx1.23\times10^{-3}$, and typical values
$x\simeq1$, $\alpha\simeq1$, one expects a dipole of amplitude $d_\kin\simeq5\times10^{-3}$, 
that is, we expect a $0.5\%$ enhancement in the number counts towards the boost apex. Since this is a small
effect, a precise measurement of it requires good sky coverage and number density.

In practice, Ellis and Baldwin's  test will  be contaminated by the gravitational
clustering of sources. Radio sources are tracers of large-scale structure, which,
when integrated along the line of sight, produces an intrinsic anisotropy in the number counts. 
Thus, in addition to the kinematic
dipole~\eqref{eqn:d-kin}, we also have a \emph{clustering dipole}, $\dint$.
The observed dipole is thus the combination of the kinematic and clustering dipole.
Unlike $\dkin$, modelling $\dint$ from first principles does require cosmological assumptions.
An interpretation of any dipole measurement needs to consider the impact of source clustering,
as the amplitude of $\dint$ may be significant if the source population is relatively local.
In any case, the expected amplitude $d_\clu$ and its probability distribution, can be predicted from theory, as we show
in this work.

The measurement of the radio dipole (and other matter dipoles) has seen a range
of results from a variety of methods and data.
Using radio sources from the NRAO VLA Sky Survey \citep[NVSS;][]{NVSS}
and a low-order spherical-harmonic fit, \citet{2002Natur.416..150B} reported one
of the first measurements of the radio dipole, finding broad agreement with the CMB
expectation (though with moderately large uncertainties).
More recent measurements \citep{2011ApJ...742L..23S,2012MNRAS.427.1994G,2013A&A...555A.117R,2017MNRAS.471.1045C,2018JCAP...04..031B}
have called on other surveys, including 2MASS, 2MRS,
BATSE gamma-ray bursts, SUMSS \citep{SUMSS}, WENSS \citep{WENSS} and TGSS;
a detailed summary can be found in \citet*{2021A&A...653A...9S}. 
These results provide further support for a larger-than-expected dipole
broadly aligned with CMB.
(Though see \citet{Darling:2022jxt} who finds agreement with the CMB
in both amplitude and direction using data
from VLASS~\citep{2020PASP..132c5001L} and RACS~\citep{2020PASA...37...48M}
catalogues).

Recently, \citet[][hereafter S21]{2021ApJ...908L..51S} provided one of the strongest
challenges to the kinematic interpretation. As with previous results (with
a direction consistent with the CMB) the amplitude was found to be
discrepant by a factor of two (i.e.~larger than expected).
On the basis of simulations, the null hypothesis (kinematic origin) was rejected at a high statistical significance of $4.9\sigma$.
This determination was enabled by a large sample of quasars selected from
the CatWISE2020
catalogue~\citep{CatWISE2020} based on the all-sky
Wide-field Infrared Survey Explorer \citep[WISE;][]{WISE}.
After applying a conservative Galactic plane cut,
this sample consisted of 1.36 million quasars covering about $50\%$ of the sky. With a favourable redshift distribution mitigating against contamination from a clustering dipole, the CatWISE quasar
sample is ideal for carrying out the Ellis and Baldwin test.

Given the implications of such a discrepancy that challenge the
basis of the standard model, it is worth reanalysing the sample of S21.
We will however take a different statistical approach.
Using a novel likelihood for the observed counts on the sky, we will
present a Bayesian analysis of the source-count map for several
parametric models.
In brief, in reaching the same qualitative conclusions as S21 we demonstrate that
the anomalous dipole amplitude is robust to methodological differences
of analysis. To assess the tension with \LCDM, we
give a first-principles comparison between our posterior
probability distribution and the theoretical prior, refining
the {\it a priori} predictions for the CatWISE dipole. In the process,
we will investigate a number of systematics (e.g.~masking, contamination
from clustering, shot noise, source evolution) on the likely values of the amplitude,
and the extent to which the discrepancy can be resolved.
The layout of this paper is as follows.
In Section~\ref{sec:approach} we outline the data sample and our approach, with the results
of our analysis presented in Section~\ref{sec:results}.
In Section~\ref{sec:pdf} we 
compute the \LCDM~prior for the probability distribution of the dipole's amplitude in the
realistic case of partial-sky coverage. We present our conclusions in
Section~\ref{sec:conclusions}. A number of appendices present the
details of the theoretical analysis undertaken in this paper.

\section{Bayesian analysis}\label{sec:approach}

\subsection{CatWISE quasar sample}
\label{sec:quasar} 
For the purposes of this study, we employ an identical sample
of quasars as presented by S21~\citep{secrest_nathan_2021_4448512}.
This is a flux-limited, all-sky sample consisting of 1,355,352
quasars selected from the CatWISE2020 catalogue~\citep{CatWISE2020},
a dataset derived from the infrared survey WISE~\citep{WISE} and NEOWISE~\citep{NEOWISE},
predominantly at $\SI{3.4}{\micro\meter}$ ({\it W1}) and $\SI{4.6}{\micro\meter}$ ({\it W2}).
The S21 sample was obtained from a magnitude cut $9<\text{\it W1}<16.4$,
with AGN selected using the colour criterion
$\text{\it W1}-\text{\it W2}\geq0.8$~\citep{2012ApJ...753...30S};
masking was applied to the Galactic plane ($|b|<\ang{30}$),
bright sources, poor-quality photometry, and image artefacts.
The spectral indices $\alpha$ in {\it W1} band were measured for each source
by power-law fits using both bands.
In this sample, $99\%$ of sources are at $z>0.1$, while $61\%$
are at $z>1$; the mean redshift is 1.2.

\begin{figure}
    \centering
	\includegraphics[width=\columnwidth]{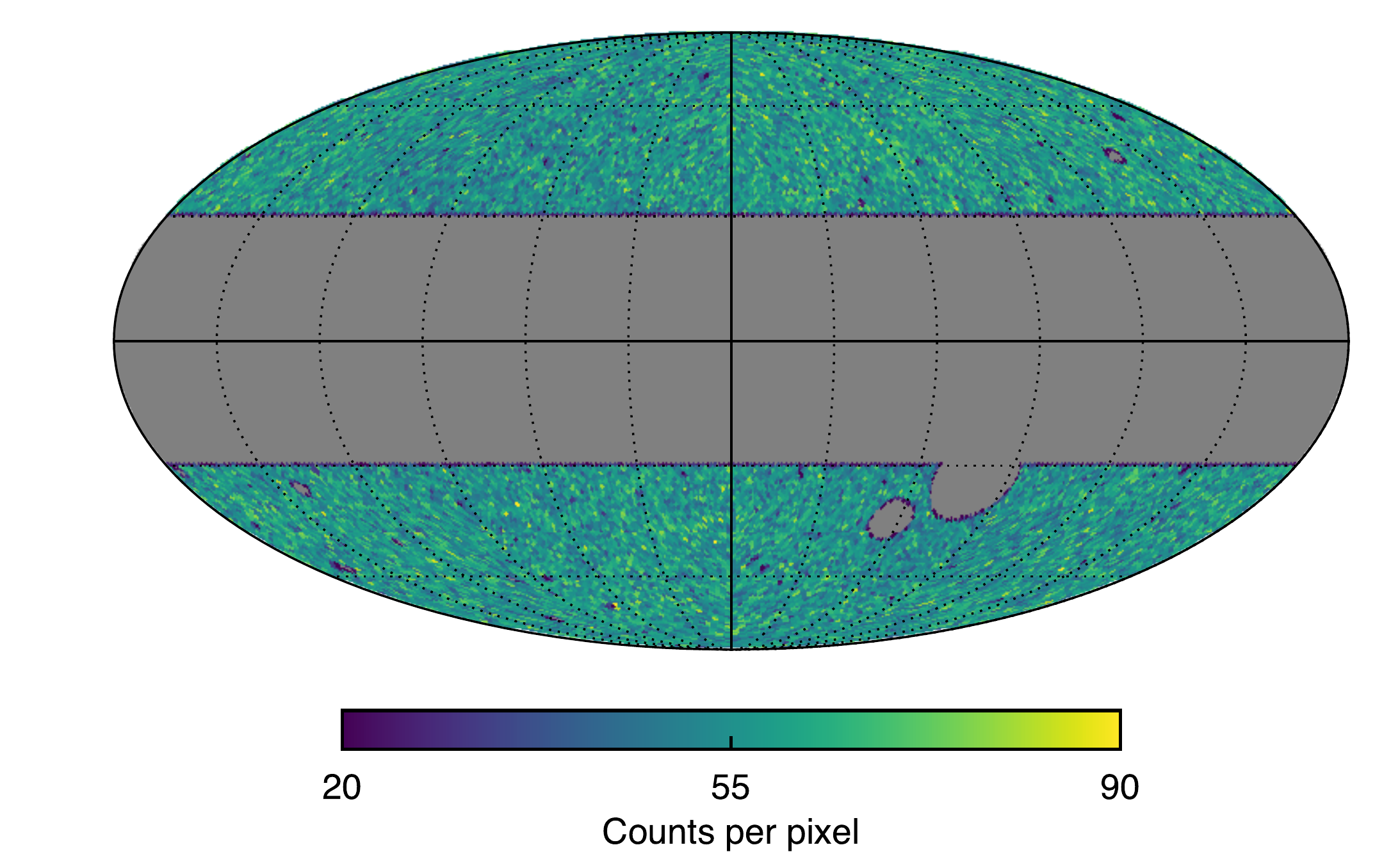}
    \caption{Mollweide map ($N_\mrm{side}=64$) of the number-count
    density of the S21 quasar sample.
    Here the Galactic-plane cut removes $|b|<\ang{30}$. The mean number
    count is 57.4 per unmasked pixel, corresponding to a surface density of 69.2 quasars per
    square degree.}
    \label{fig:map}
\end{figure}

In determining the presence of a dipole structure in the
resultant quasar sample, we first binned the counts over the
sky to determine their density. To achieve this, {\tt healpy} \citep[a Python
implementation of \healpix;][]{2005ApJ...622..759G,Zonca2019}
was used to determine equal area pixels over the sky. 
For this study, $N_\mrm{side}=64$, which corresponds to a total number of pixels of $49,152$, each 
with an area of 0.83 square degrees.
The total number of unmasked pixels is $\npix=23,972$, and 
the pixelized map analysed in this work is shown in Fig.~\ref{fig:map}.

\subsection{Parametric model and likelihood}\label{sec:definition}
The expected number density of quasars in direction $\n$ is given to a good approximation
by the dipolar modulation (i.e. the sum of the monopole and dipole moments)
\begin{equation}
    \frac{\dif N}{\dif\Omega}(\n) = \bar{N} \big(1 + \mbf{d}\cdot\n\big),
    \label{eqn:dipole}
\end{equation}
where $\n$ is the direction of observation, $\bar{N}$ is the mean number density on the sky,
and $\mbf{d}\cdot\n$ is the dipole. This equation forms the basis of
our analysis.
In practice, there are a number of systematics to be considered.
In particular, S21 identified a scanning pattern bias in
the CatWISE2020 
sample that varies linearly away from the ecliptic equator, such that 
there is a decrease of source density of $\SI{\sim7}{\%}$ at the ecliptic poles.
To allow for this possibility we introduce a biasing term of the
form
\begin{equation}
f_\mrm{ecl}(\n) \equiv 1 - \Upsilon_\mrm{ecl} \sp c_\mrm{ecl}\sp |b_\mrm{ecl}(\n)|,
\label{eqn:eclipticbias}
\end{equation}
where the slope $c_\mrm{ecl}=7.4\times10^{-4}$ was determined by S21 through
a linear fit, and $b_\mrm{ecl}(\n)$ is the ecliptic latitude.
We introduced $\Upsilon_\mrm{ecl}$, a nuisance parameter which exactly recovers the S21 correction 
when $\Upsilon_\mrm{ecl}=1$, 
and corresponds to no ecliptic bias when $\Upsilon_\mrm{ecl}=0$.
\begin{figure*}
    \centering
    \includegraphics[width=1.8\columnwidth]{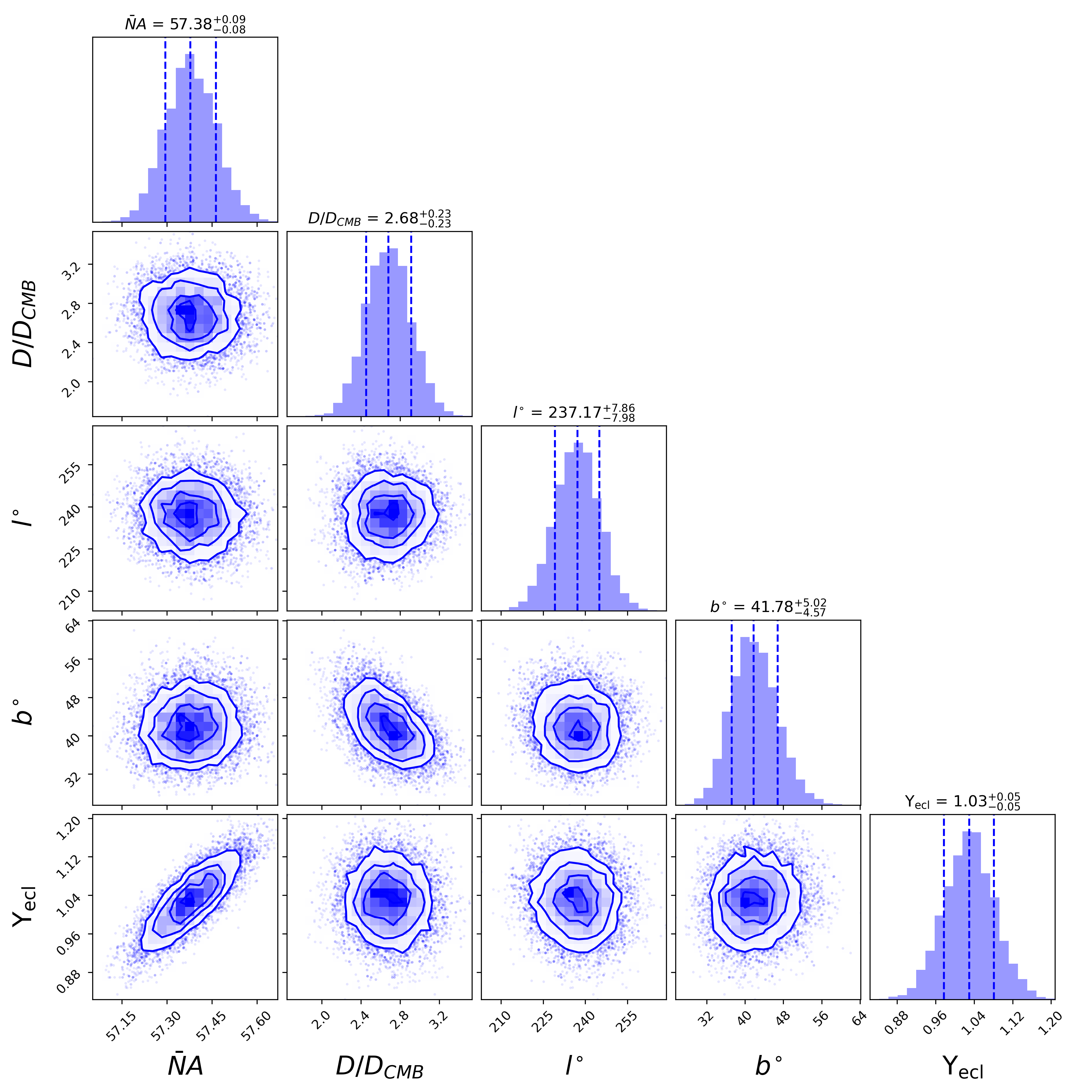}
    \caption{Corner plot for the results of the MCMC exploration of the posterior distribution.
      Dashed vertical lines indicate the mean and $68\%$ confidence limits (as quoted).
      Here $\bar{N}A\equiv\bar{N}\Omega_\mrm{pix}$ is the mean number count per pixel, $D/D_\text{CMB}$ is the dipole amplitude normalised relative
      to the CMB expectation, and $\Upsilon_\mrm{ecl}$ is the
      ecliptic bias parameter; the
      Galactic coordinates $(l,b)$ specify the direction of the dipole $\mbf{d}$ on the sky. Note that $D_\text{CMB} = 0.0072$, as determined from the
      CMB velocity~\eqref{eqn:vel-hel} and the sample's
      mean values $\bar\alpha=1.26$ and $\bar{x}=1.7$.
    The contours correspond to  0.5, 1, 1.5, and $2\sigma$, encompassing
    11.8\%, 39.3\%, 67.5\%, and 86.4\% of the samples, respectively \citep{corner}.
    }
    \label{fig:corner}
\end{figure*}
With the inclusion of the ecliptic bias, the expected number  of 
quasars in a \healpix\ pixel at $\n$ is given by the parametric model
\begin{equation}
\lambda(\n) = \Omega_\mrm{pix}\sp f_\mrm{ecl}(\n)\frac{\dif N}{\dif\Omega}(\n) ,
\label{eqn:totallambda}
\end{equation}
where $\Omega_\mrm{pix}$ is the pixel area.

The counting statistics within
each pixel motivates a likelihood of Poisson form, with
mean varying from pixel to pixel due to the expected dipolar
modulation.
Hence, given the expected
number of quasars $\lambda_i=\lambda(\n_i)$ in the $i$th pixel, the probability of observing $N_i$ quasars
in that pixel is
\begin{equation}
    p(N_i\mid\theta) = \frac{\lambda_i\Zu{N_i}\sp \rme\Zu{-\lambda_i}}{N_i !}.
    \label{eqn:poisson}
\end{equation}
The likelihood of obtaining number counts $N_1,N_2,\ldots,N_\npix$ (the data $\mathscr{D}$)
is therefore
\begin{equation}
    \mathcal{L}(\theta;\mathscr{D})
        = \prod_{i=1}^{\npix} p(N_i\mid\theta),
    \label{eqn:likelihood}
\end{equation}
where $\theta$ is the set of free parameters, and the product
is taken over all $\npix$ unmasked pixels.
In this work $\theta=\{\bar{N},D,l,b,\Upsilon_\mrm{ecl}\}$, where $D\equiv|\mbf{d}|$ is the
dipole amplitude, $l,b$ are the Galactic coordinates of $\mbf{d}$ (and $\bar{N}$ and $\Upsilon_\mrm{ecl}$ were described above).

One advantage of our Bayesian approach over
other approaches often used is uncertainty quantification. This is a routine task once the posteriors are sampled, and yields an internal measure of the uncertainty (no simulations are required). Furthermore, the posterior can be directly
compared with the distribution we expect from theory
(Section~\ref{sec:pdf}).
Indeed, previous estimates of the dipole are largely based on constructing estimators; see, e.g.~\citet{2021A&A...653A...9S} and references
therein.
However, in the situation where all cells have high occupancy number, the 
Poisson likelihood converges to a Gaussian, with both mean and variance equal to $\lambda_i$. In this limit we can derive from equation~\eqref{eqn:poisson} the maximum-likelihood estimator
of the dipole, which is
equivalent (up to some normalisation) 
to the quadratic estimator commonly used.
(With $N_\mrm{side}=64$ we have about $57$ sources per pixel, so we
are well within this regime.)
Although we will only extract the monopole and dipole, we note
that the likelihood~\eqref{eqn:likelihood} also retains complete information in the pixelized map, including of higher multipoles.

\subsection{Hypotheses considered}
For the purposes of this study, several differing hypotheses were considered and compared.
\begin{itemize}
\item $M_0$: Taken to be the null hypothesis, this assumes that there is no dipole or ecliptic bias in the CatWISE sample, and only has a single free parameter, $\bar{N}$. $D$ and $\Upsilon_\mrm{ecl}$ are fixed at zero.
\item $M_1$: The dipole's amplitude is precisely that of the CMB
but its direction may differ from that of the CMB. No ecliptic bias is assumed, so $\Upsilon_\mrm{ecl}=0$.
\item $M_2$: This assumes that a dipole is present and has the same magnitude and direction as that observed in the CMB. For this, the amplitude is fixed to $D_\text{CMB} = 0.0072$, whilst the direction is 
fixed at $(l,b) =  (264\dotdeg021,48\dotdeg253)$.
Again, $\bar{N}$ is treated as a free parameter and no ecliptic bias is assumed ($\Upsilon_\mrm{ecl}$).
\item $M_3$: Again, this assumes that a dipole is present, and is aligned with the CMB. However, the amplitude of the dipole term, $D$ is treated as a free parameter, as is $\bar{N}$. The direction is 
fixed at $(l,b) =  (264\dotdeg021,48\dotdeg253)$, and no ecliptic bias is assumed ($\Upsilon_\mrm{ecl}=0$). 
\item $M_4$: Here, the presence of a dipole is assumed, but its amplitude and direction, $D, l$ and $b$ are taken as as free parameters, as is $\bar{N}$. The ecliptic bias, $\Upsilon_\mrm{ecl}$, is again assumed to be zero. 
\end{itemize}
Models $M_1$, $M_2$, $M_3$, $M_4$ were repeated (as $M_5$, $M_6$, $M_7$, $M_8$, respectively) but with the ecliptic bias $\Upsilon_\mrm{ecl}$ treated as a free parameter.
These models are summarised in Table~\ref{tab:evidence}.

\begin{table}
\caption{Marginal likelihoods for the various models considered. Note that $Z$ is increasing from $M_0$ to $M_8$.}
\begin{tabular}{l|l|l}
\hline
Model  & Description & $\ln(Z)$ \\
\hline
$M_0$  & Null (no dipole) & -87707.67 \\
$M_1$ & Amplitude fixed to CMB; no bias & -87654.43 \\
$M_2$ & Amplitude and direction fixed to CMB; no bias & -87652.04 \\
$M_3$ & Direction fixed to CMB; no bias & -87625.43 \\
$M_4$ & All parameters free, except no bias & -87624.18 \\
$M_5$ & Amplitude fixed to CMB & -87473.38 \\
$M_6$ & Amplitude and direction fixed to CMB & -87472.77 \\
$M_7$ & Direction fixed to CMB & -87445.27 \\
$M_8$ & All parameters free & -87444.17 \\
\hline
\end{tabular}
\label{tab:evidence}
\end{table}

\begin{figure}
    \centering
	\includegraphics[width=1.\columnwidth]{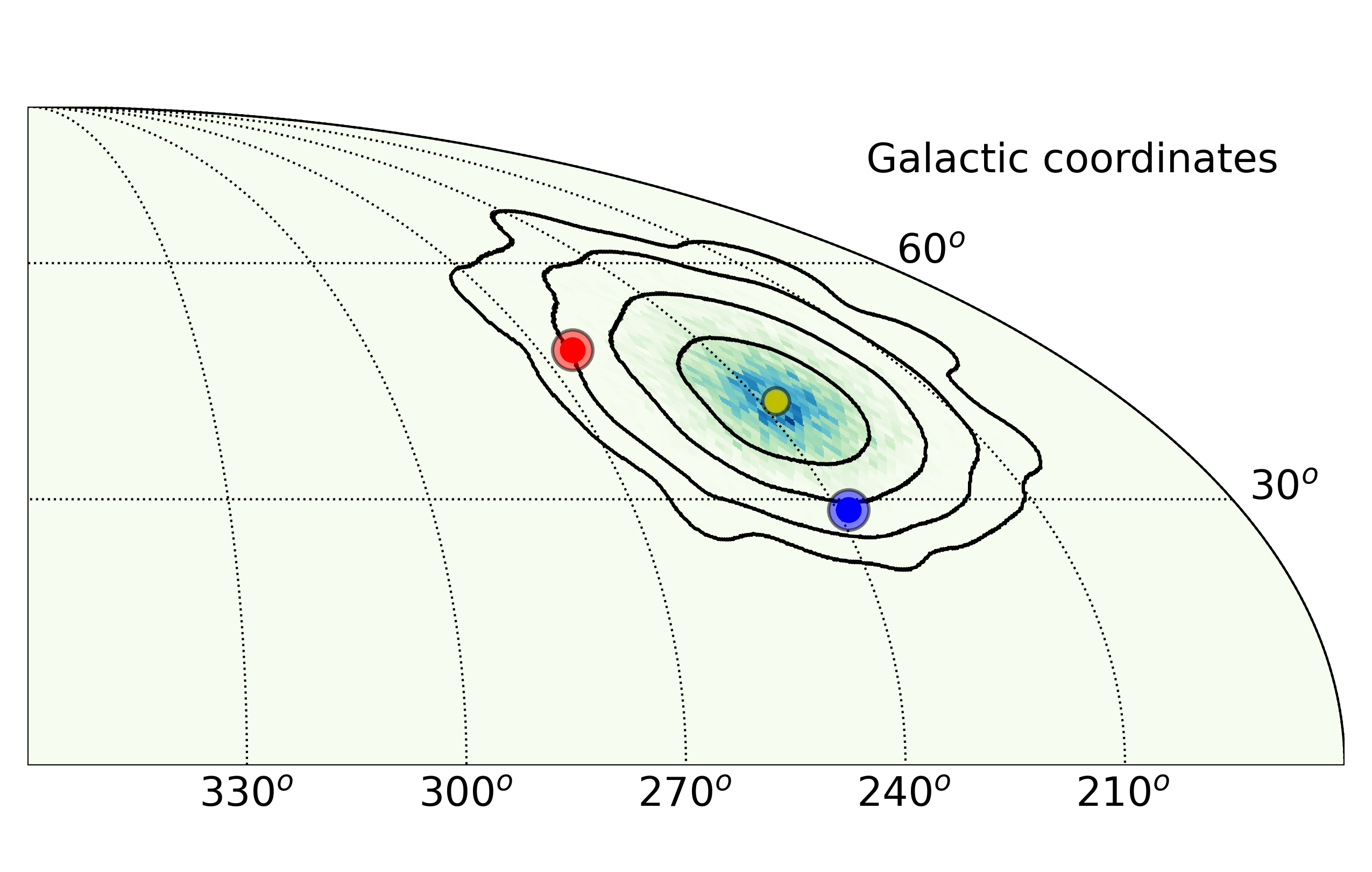} 
    \caption{The on-sky posterior distribution of the direction of the dipole from the MCMC exploration, with the colour-coding corresponding to a distribution normalised to unity. The contours are derived from smoothing the underlying distribution with a Gaussian kernel with smoothing scale $\sigma=\ang{3}$, with each contour encompassing $68\%$, $95.5\%$, $99.7\%$ and $99.99\%$ of the probability. The yellow-filled circle indicates the point of the best-fit values, whereas the red-filled circle denotes the direction of the dipole as determined by \citet{2020A&A...641A...3P}.
    The blue-filled circle represents the dipole direction as determined by S21. 
    }
    \label{fig:galactic}
\end{figure}

\subsection{Posterior exploration\label{sec:results}}
With given model $M_i$ and data $\mathscr{D}$,
the posterior probability distribution of obtaining the
set of parameters $\theta$ is by Bayes' theorem
\be
p(\theta\mid\mathscr{D})=\mathcal{L}(\theta;\mathscr{D})\sp p(\theta)/Z,
\ee
where $p(\theta)$ is the prior and
$Z\equiv\int\dif\theta\sp \mathcal{L}(\theta;\mathscr{D})\sp p(\theta)$ is
the marginal likelihood, or Bayesian  evidence.
We explored the posterior 
using the DNest4 sampler~\citep[Diffusive Nested Sampling;][]{JSSv086i07}, allowing a determination of the posterior distributions, but also providing the marginal likelihood. This allows an effective comparison of the competing hypotheses based on how well they predicted the data,
marginalised over the parameter space of each model. 

For the purposes of this study, the following priors were adopted.
For the mean number count $\bar{N}\Omega_\mrm{pix}$ a uniform distribution between $40$ and $60$ was chosen. A uniform distribution was also adopted for the scaled amplitude of the dipole, $D / D_\text{CMB}$, between $0$ and $10$;
as noted in S21, the expected amplitude of dipole is $D_\text{CMB} = 0.0072$ (for the sample's mean values
$\bar\alpha=1.26$ and $\bar{x}=1.7$).
The direction of the dipole in $l$ and $b$ were chosen to give a uniform distribution over the entire sphere, for $\ang{0} < l < \ang{360}$ and $\ang{-90} < b < \ang{90}$. A uniform prior was adopted for the
ecliptic bias, $-2 < \Upsilon_\mrm{ecl} < 2 $.

The resulting Bayesian evidences for the various hypotheses are presented in Table~\ref{tab:evidence}. There are two immediate points to take from these. First, the null hypothesis of there being no dipole is overwhelmingly disfavoured, demonstrating that a non-homogeneous signature is present in the distribution of quasars. Second, the models correcting for the ecliptic bias are strongly favoured over those models that do not consider this component. Hence, in the following we will focus upon models $M_6$, $M_7$ and $M_8$.

It is clear that $M_6$ -- which assumes a dipole with a fixed amplitude and direction to match the expectations of the kinematic dipole -- is highly disfavoured when compared to $M_7$ and $M_8$.
For $M_7$, where the amplitude of the dipole is free to vary, we find $D / D_\text{CMB} = 2.46 \pm 0.18$, 
demonstrating that, as with S21, the amplitude of the quasar dipole is significantly different 
to the expectation from the CMB.
Intriguingly, the ratio of the Bayesian evidences for $M_7$ and $M_8$ is only 3, in favour of $M_8$. Therefore, we find only mild evidence that the dipole direction differs from the CMB direction. 
For completeness, Fig.~\ref{fig:corner} presents the corner plot of the posterior distributions for $M_8$. 
As well as showing the evidence for the ecliptic biasing as identified by S21, with $\Upsilon_\mrm{ecl} = 1.03 \pm 0.05$, again a significantly larger dipole amplitude, with $D / D_\text{CMB} = 2.68 \pm 0.23$, is identified. From the posterior distribution we compute $P(D>D_\text{CMB}\mid\mathscr{D})=\int^\infty_{D_\text{CMB}}\dif D\ssp p(D\mid\mathscr{D})$, 
the probability that the CatWISE dipole amplitude exceeds the CMB value, all other parameters
marginalised over. We find $P(D>D_\text{CMB}\mid\mathscr{D})=0.9999999888$,
corresponding to a statistical significance of $5.7\sigma$.

As for the direction of the dipole we find
$(l,b) = (237\dotdeg2^{+7\dotdeg9}_{-8\dotdeg0} , 41\dotdeg8\pm5\dotdeg0)$. 
The posterior distribution of the dipole directions on the sky is presented in Fig.~\ref{fig:galactic}, with contours indicating the 
 $68\%$, $95.5\%$, $99.7\%$ and $99.99\%$ credible regions. The yellow-filled circle indicates the point of the best-fit values, whereas the red-filled circle denotes the direction of the dipole as determined by \citet{2020A&A...641A...3P}. The blue-filled circle represents the dipole direction as determined by S21. As demonstrated by the ratio of the Bayesian evidences for $M_7$ and $M_8$, this direction is only mildly favoured 
 over the quasar dipole being aligned with that detected in the CMB.

It is important to note that Bayesian model selection calculations, such as those presented 
in this paper, can be affected by the choice of priors. Particularly, if we had used a narrower
prior for the direction of the quasar dipole (i.e.~if we had
expected it to be {\em close} to the CMB direction), we would have
found stronger evidence for a direction difference. On the other hand,
if we had used a wider prior for parameters like
$\bar{N}A$ that have the same meaning across models,
this would decrease the evidence values across the board, but leave conclusions about
the dipole unchanged. However, given that we should not necessarily expect {\it a priori} the quasar 
and CMB dipole to be aligned, we are justified in using the large prior range adopted.

\section{A priori predictions for the dipole amplitude}\label{sec:pdf}
In light of renewed interest in dipole anomalies it is worth
revisiting the predictions of \LCDM~\citep[e.g.][]{2012MNRAS.427.1994G,2014A&A...565A.111R,2016JCAP...03..062T,2019MNRAS.486.1350B},
without invoking the stringent CMB prior~\eqref{eqn:vel-hel},
which assumes a purely kinematic origin of the dipole.
Here we are primarily interested in the distribution of values
of the amplitude. In order to obtain this we will need the
probability distribution of the dipoles themselves.
The key assumption we will make is that the kinematic and
clustering dipoles are Gaussian distributed. This is justified
in the case of the clustering dipole given that it is to
a large degree sourced by matter fluctuations well described by
linear theory. The kinematic dipole, on the other hand, is harder
to justify. The local velocity is the result of several contributions,
including the motion of the Earth around the Sun; the Sun around
the Milky Way; the virial motion of the Milky Way within the
Local Group, etc. Assigning a distribution
to each of these motions is beyond what we can hope to achieve in
this comparison. Here we will take a decidedly less fine-grained approach.
We will assume that the local velocity admits a decomposition
into a linear and nonlinear component, as described by the halo model.
The linear contribution, due to infall onto large-scale structure, can
then be described by linear theory and is Gaussian in nature. The nonlinear
contribution is generically ascribed to virial motions within a spherical halo
described as an isothermal sphere. In this case the motions are Gaussian for
a given halo mass, in reasonable agreement with simulations~\citep{2001MNRAS.322..901S}.
However, it is important to note that the probability distribution,
averaged over all halo masses, is \emph{not} Gaussian, exhibiting 
larger-than-Gaussian tail probabilities due to virial
motions~\citep{Cooray:2002dia}. We will return to this point later in
Section~\ref{sec:numerical-results}.

Now since the dipoles are Gaussian, the amplitude is
described by the Maxwellian. This is in general a long-tailed distribution
with large values of $D$ not precluded. However, this is strictly only true
for full-sky coverage; for partial-sky coverage the distribution is no longer
Maxwellian due to the loss of isotropy. As we will show,
in the partial-sky case there are non-trivial covariances
between different multipole moments (e.g.~in the clustering
statistics), leading generically to a leakage of power from
higher multiples (smaller angular scales) into the dipole,
changing the theoretical expectations of the amplitude.
With $50\%$ of the sky removed, the impact of masking
is sizeable.
In this section we will thus compare the posterior $p(D\mid\mathscr{D})$
obtained in the previous section, with the theoretical prior $p(\Dtot)$,
the probability distribution of the total dipole amplitude
$\Dtot=|\mbf{d}|$ according to \LCDM. Since the computation of
$p(\Dtot)$ is somewhat involved we present here only the main results,
relegating the technical details to appendices.

\subsection{Dipole statistics}\label{sec:marge}
The dipole anisotropy has a direction and amplitude and it is convenient to represent it as a three-dimensional Cartesian vector $\mbf{d}$
such that $\mbf{d}\cdot\n=\sum_{m=-1}^1 a\Z{1m}Y_{1m}(\n)$, where
$Y_{1m}(\n)$ are the $\ell=1$ spherical harmonics~\citep*{2004PhRvD..70d3515C}.
Since the dipole is estimated on the masked sky, in order to
compare with theoretical predictions we will need to take
this into account.
We will therefore write the number-count fluctuations as
\begin{equation}\label{eqn:Delta-N}
    \Delta N(\n)/{N}
        =\mathcal{W}(\n)\big({\dkin}\cdot\n + \dqso(\n)\big)
        =\mathcal{W}(\n)(\mbf{d}\cdot\n+\cdots),
\end{equation}
where $\mathcal{W}(\n)$ is the mask and $\mbf{d}=\dkin+\dint$.
Here the ellipsis represent higher multipoles ($\ell\geq2$) of the intrinsic fluctuations $\dqso(\n)=\sum_{\ell m}a\Z{\ell m}Y_{\ell m}(\n)$.
For \LCDM\ these multipoles are subdominant to the dipole and may therefore be ignored.
The underlying kinematic and clustering dipoles (on the unmasked sky) are uniquely given in terms of their harmonic coefficients by
\begin{subequations}\label{eqn:dipole-vecs}
\bea
    \dint
        &=\sqrt{\frac{3}{4\pi}}
        \big(-\sqrt{\sp2}\sp\Re(a\Z{11}),\ssp \sqrt{\sp2}\sp\Im(a\Z{11}),\ssp a\Z{10}\,\big)\Zu\T, 
        \label{eqn:d-int} \\[1pt]
    \dkin
        &=\sqrt{\frac{3}{4\pi}}
        \big(-\sqrt{\sp2}\sp\Re(b\Z{11}),\ssp \sqrt{\sp2}\sp\Im(b\Z{11}),\ssp b\Z{10}\,\big)\Zu\T,
        \label{eqn:b-LM}
\eea
\end{subequations}
with
\begin{subequations}
\bea
    a\Z{\ell m}
        &= \int\dif^2\n~Y_{\ell m}^*(\n)\ssp\dqso(\n), \\
    b\Z{\ell m}
        &= \int\dif^2\n~Y_{\ell m}^*(\n)\ssp (\dkin\cdot\n). 
\eea
\end{subequations}
Note that since $\dkin\cdot\n$ is a pure dipole we have that only for $\ell=1$ are the
$b_{\ell m}$ non-zero. This is not the case for $\dqso$, however.

In general, the fluctuation $\dqso(\n)$ results from a number of effects, including
redshift-space distortions~\citep{1987MNRAS.227....1K}, Doppler effects, gravitational lensing~\citep{1984ApJ...284....1T,2022MNRAS.510.3098M}, gravitational redshift,
and relativistic corrections~\citep*{Yoo:2009au,2011PhRvD..84f3505B,2011PhRvD..84d3516C}. The complete expression for $\dqso(\n)$ can be found in, e.g.\ \citet[appendix A]{CLASSgal}; we compute it
using a modified version of {\sc class}~\citep*{CLASS}.
Since the distribution of the amplitude is largely insensitive to the bias,
here we simply fix the bias at the mean redshift using the
\citet{2005MNRAS.356..415C} quasar bias parametrisation.
Here since $\bar{z}=1.2$ we have $\bqso=2$.

As for the individual statistics of $\dkin$ and $\dint$
these are
given by trivariate Gaussians. The joint distribution
is also
a Gaussian, containing non-trivial correlations between $\dkin$ and $\dint$.
This is because the kinematic dipole is in part sourced
from our being pulled by the surrounding
matter distribution, implying that there is necessarily a clustering contribution.
Since these two contributions arise from the same large-scale structure, and are linearly
related to the matter distribution $\delta(\x)$, their
joint statistics must also be Gaussian.
We will therefore take $(\dkin,\dint)\sim\calN(\bm0,\cov)$, with the $6\times6$
covariance matrix
\bea\label{eqn:cov-6x6}
    \cov
    &=
        \begin{pmatrix}
            \,\big\langle\dkin\Zu{\phantom\T}\dkin\Zu\T\big\rangle &
            \big\langle\dkin\Zu{\phantom\T}\dint\Zu\T\big\rangle\: \\[4pt]
            \,\big\langle\dkin\Zu{\phantom\T}\dint\Zu\T\big\rangle &
            \big\langle\dint\Zu{\phantom\T}\dint\Zu\T\big\rangle\:
        \end{pmatrix}. 
\eea
The structure of this covariance will depend on whether
we are working on the full sky or the cut sky.
For full-sky coverage each block matrix is proportional
to the identity matrix (on account of isotropy), but in general
it is more complex.

Note on account of equation~\eqref{eqn:dipole-vecs} that the
harmonic coefficients are linearly related to the dipole vectors, so are also Gaussian distributed. We will switch to
the harmonic description (e.g.~in Appendix~\ref{app:galactic-cut}),
which is more convenient when incorporating a masked.

\subsection{Probability distribution of the amplitude}\label{sec:p-Dtot-cutsky}
We are of course interested in the amplitude of the dipole, $D=|\dkin+\dint|$, and in particular its distribution $p(\Dtot)$,
the theoretical \LCDM~prior.
It can be computed as follows.
Since the amplitude depends on the individual dipoles, $D\equiv|\dkin+\dint|$, we have by the chain rule
\be
    p(\Dtot)
        =\int \dif^3\dkin\int\dif^3\dint\
            p(\Dtot\mid\dkin,\dint)\ssp
            p(\dkin,\dint),
    \label{eqn:p-Dtot-int}
\ee
i.e.~we marginalise over each dipole.
Here $p(\dkin,\dkin)=\calN(\bm0,\cov)$, whereas the relation between
the amplitude and the dipoles is fixed, so $p(\Dtot\mid\dkin,\dint)=\delta_{\rm D}\big(\Dtot-|\dkin+\dint|\big)$.

In the case of full-sky coverage the statistics of the dipoles are
isotropic, and the marginalisation~\eqref{eqn:p-Dtot-int} yields the Maxwell--Boltzmann distribution,
as is well known. In the more realistic case of partial-sky coverage the PDF is
no longer Maxwellian; in general, the variances of each
component are not identical.
However, with the Galactic plane cut of S21
the marginalisation can still be done. We find
(see Appendix~\ref{app:pD_derivation} for details)
\be\label{eqn:p-Dtot-cutsky}
\begin{split}
    p(\Dtot)\,\dif\Dtot
        &=(1-\varepsilon)\sp\mathcal{F}\Big(\sqrt{|\varepsilon|}\ssp\frac{\Dtot}{\sqrt2\Delta_z}\ssp\Big)\sp
        f_\mrm{MB}\Big(\frac{\Dtot}{\sqrt2\Delta_z}\Big)
        \frac{\dif\Dtot}{\sqrt2\Delta_z},
\end{split}
\ee
where $\mathcal{F}(x)\equiv F(x)/x$, in which
$F(x)\equiv\rme^{-x^2}\int^x_0\dif t\,\rme^{t^2}$ is the Dawson function;
we have introduced $\varepsilon\equiv1-\Delta^2_z/\Delta^2_x$, which parametrises the
`eccentricity' between the dispersions along the $z$-axis and the $x$-axis (or, equivalently, the $y$-axis) as induced by our azimuthal sky cut; and
$f_\mrm{MB}(x)$ has the Maxwell--Boltzmann form,
$f_\mrm{MB}(x)\dif x={4}/{\sqrt{\pi}}\,x^2 \rme^{-x^2}\dif x$.
Here we have
\begin{subequations}
\bea
    \Delta^2_x
        &=\sigma^2_{\kin_x\kin_x} + 2\sp\sigma^2_{\kin_x\clu_x} + \sigma^2_{\clu_x\clu_x}, \label{eqn:Delta_x} \\
    \Delta^2_z
        &=\sigma^2_{\kin_z\kin_z} + 2\sp\sigma^2_{\kin_z\clu_z} + \sigma^2_{\clu_z\clu_z}, \label{eqn:Delta_z}
\eea
\end{subequations}
which give the dispersions for the $x$- and $z$-components of the total dipole $\mbf{d}=\dkin+\dint$ (with the
dispersion of the $y$-component the same as that of the $x$-component).%
\footnote{Although $\dint$ here is not strictly a
velocity, it is related to the underlying density field
$\delta(\x)$ (see Appendix~\ref{app:b-1M-derivation})
so we will also consider it a `velocity', with some 
associated dispersion.}
Notice that this PDF~\eqref{eqn:p-Dtot-cutsky} is
similar to the Maxwellian,
save for the first two factors, which
depend on the asymmetry of the cut.
The overall effect of these factors suppresses
the tail of the Maxwellian.
The modified PDF has mean
$\langle\Dtot\rangle
=\sqrt{2/\pi}\ssp\big[\Delta_z+\Delta_x\sp\mrm{arcsinh}(\sqrt{|\varepsilon|})/\sqrt{|\varepsilon|}\ssp\big]$,
and r.m.s.~value $\sqrt{\langle\Dtot^2\rangle}=(3-\varepsilon)^{1/2}\Delta_x$.%
\footnote{In the case of full-sky coverage
we recover the Maxwellian, which has identical
dispersion along each axis,
$\Delta^2_x=\Delta^2_y=\Delta^2_z\equiv\Delta^2$, with
$\Delta^2=\sigkin^2 + 2\sp\rho\sp\sigkin\sp\sigint + \sigint^2$. We then have
for the r.m.s.~amplitude $\sqrt{\langle\Dtot^2\rangle}=\sqrt3\sp\Delta$,
and mean $\langle\Dtot\rangle=\sqrt{8/\pi}\sp\Delta\approx0.92\sqrt{\langle\Dtot^2\rangle}$.}
We find that the mean value $\langle D\rangle$ on the cut sky is
half that on the full sky.

\subsubsection{\LCDM~considerations}\label{sec:predictions}
We now evaluate the probability distribution~\eqref{eqn:p-Dtot-cutsky}, plugging in the detailed
predictions of $\Lambda$CDM for $\Delta_x^2$ and $\Delta_z^2$.
In general, for two tracers of large-scale structure, labelled $\mrm{X}$ and $\mrm{Y}$, these dispersions are given in terms
of the angular power spectra by
\be
    C^{\mrm{XY}}_\ell
        =4\pi\int^\infty_0\frac{\dif k}{k}\,\Delta_\ell^\mrm{X}(k)\,\Delta_\ell^\mrm{Y}(k)\,\PR(k),
    \label{eqn:Cab-ell}
\ee
where $\Delta_\ell^\mrm{X}(k)$ is the harmonic transfer function
of $\mrm{X}$, and $\PR(k)$ is the (dimensionless) power spectrum of primordial curvature
perturbations ${\zeta}(\bk)$ defined by
$\langle{\zeta}(\bk)\ssp{\zeta}^*(\bk')\rangle=(2\pi)^3\delta_{\rm D}(\bk-\bk')\sp 2\pi^2\PR(k)/k^3$.
The analytic forms of the kinematic ($\mrm{X}=\kin$) and clustering ($\mrm{Y}=\clu$) transfer functions are
presented in Appendix~\ref{app:b-1M-derivation}. Note
that the clustering transfer function depends on the
sample's redshift distribution $p_z(z)$; this has been
estimated by S21 by cross-matching a subsample of their sources
with those in a subregion of SDSS (Stripe 82).

Since $\dkin=\ckin\ssp\bm\beta$, where $\bm\beta=\v\Z{O}/c$ and
$\ckin\equiv2+x(1+\alpha)$, to fix the statistics of the kinematic dipole we will need a model of the local velocity.
We will take this velocity to be composed of a smooth
part $\v\Z{R}$ and a stochastic part $\v_\mrm{vir}$, i.e.~$\v\Z{O}=\v\Z{R}+\v_\mrm{vir}$.
The smooth part, coherent over large scales, can be described using linear perturbation theory; on the other hand the
stochastic part, due to the virial motions of clusters, is
nonlinear in nature.
Given that $\v\Z{R}$ and $\v_\mrm{vir}$ arise from different physical
processes, on different length scales,
we will take these two types of motion to be uncorrelated, $\langle\v\Z{R}\v_\mrm{vir}^\T\rangle=\langle\v\Z{R}\rangle\langle\v_\mrm{vir}\rangle\Zu\T=0$.

The coherent part $\v\Z{R}$ is sourced by the
matter distribution,
smoothed by a spherically-symmetric window function with characteristic length $R$, i.e.\
\bea
    \v\Z{R}(\x)
        &\equiv\int\dif^3\x'\ W_R(\x-\x')\ssp \v(\x') \nonumber\\[-2pt]
        &=\int\frac{\dif^3\bk}{(2\pi)^3}\, \frac{-\im\bk}{k}\bigg(\frac{H_0 f_0}{k}\bigg)\ssp
                    W(kR)\sp\delta(\bk)\sp \rme^{-\im\bk\cdot\x}.
\eea
Here $f_0\approx\OmO\Zu{0.55}$ is the present-day growth rate, and
in the second line we have used the linearised continuity equation
to relate the velocity to the matter distribution.
In this work we adopt a spherical top-hat filter for which
$W(kR)=3j_1(kR)/(kR)$, where $R$ is
the comoving Lagrangian radius associated with mass $M$ through
$M=4\pi R^3\bar\rho_{m0}/3$.

\begin{table}
\centering
\caption{Cluster masses considered and the corresponding smoothing scales. Here $\sigma_\mrm{vir}$, $\sigma\Z{R}$,
and $\sigma_\mrm{tot}=(\sigma_\mrm{vir}^2+\sigma\Z{R}^2)^{1/2}$
  are the one-dimensional dispersions.
  }
\setlength{\tabcolsep}{2pt}
\begin{tabular}{ccccccc}
\toprule
$M$ [$h^{-1}\ssp M_\odot$] & $R$ $[h^{-1}\ssp\mrm{Mpc}]$ & $\sigma_\mrm{vir}$ $[\mrm{km\,s^{-1}}]$ & $\sigma\Z{R}$ $[\mrm{km\,s^{-1}}]$ & $\sigma\Z{\mrm{tot}}$ $[\mrm{km\,s^{-1}}]$ \\ 
\midrule
$10^{12}$ & $1.26$ & $ 49$ & $306$ & $310$ \\ 
$10^{13}$ & $2.71$ & $105$  & $304$ & $321$ \\ 
$10^{14}$ & $5.85$ & $226$  & $295$ & $372$ \\ 
$10^{15}$ & $12.6$ & $488$  & $276$ & $561$ \\ 
\bottomrule
\end{tabular}
  \label{tab:smooth}
\end{table}

The virial motion is assumed to be that of an
isothermal sphere of mass $M$~\citep{2001MNRAS.322..901S}. Then, by the virial theorem,
$\sigma_\mrm{vir}^2\propto M/R_\mrm{vir}\propto M^{2/3}$, and by isotropy
$\langle\v_\mrm{vir}^{\phantom{\T}}\v_\mrm{vir}^\T\rangle=\sigma_\mrm{vir}^2\ssp\mat{I}_3$.
We will compute the present-time velocity dispersion using
the fitting formula~\citep{1998ApJ...495...80B}
\be
    \sigma_\mrm{vir}(M)
        =\frac{1}{\sqrt3}\, 476\ssp g\Z{\sigma}\ssp\Delta_\mrm{vir}\Zu{1/6}\ssp
         \bigg(\frac{M}{10^{15}\ssp h^{-1}\sp M_\odot}\bigg)^{1/3}\, \mrm{km\,s^{-1}},
\ee
where $g_\sigma=0.9$, and $\Delta_\mrm{vir}=18\pi^2+60y-32y^2$, with $y=\OmO-1$.%

\begin{figure}
\centering
    \begin{subfigure}{\columnwidth}
        \centering
	    \includegraphics[width=\textwidth]{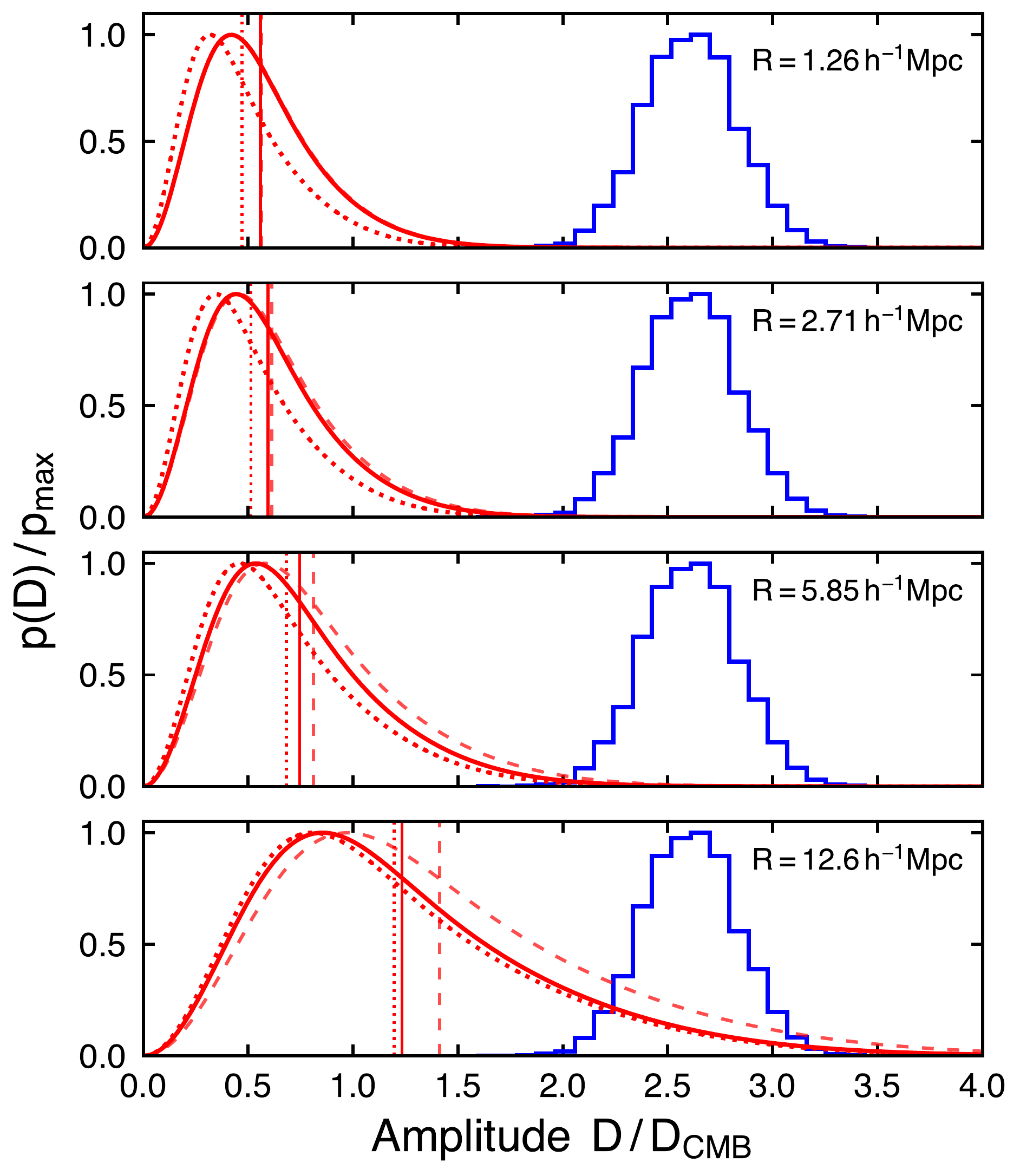} 
    \end{subfigure}
    \caption{Comparison between the theoretical prior (red) and the recovered posterior (blue; as shown in Fig.~\ref{fig:corner}) of the dipole amplitude. (Note that the posterior is based
    on a uniform prior on $D$.)
    In each subplot -- corresponding to a different smoothing
    length, or velocity dispersion -- we show the theoretical prior
    with contributions from clustering (solid curve) and without them (dotted curve); the dashed curve indicates the
    distribution when both clustering and the maximal 
    source-evolution correction are included ($\rho=1$; see equation~\eqref{eqn:d-kin-corr}).
    The vertical lines indicate the means of the respective probability densities.
    Here we have taken
    $x=1.7$
    (corresponding to a magnification bias of $s=2x/5=0.68$),
    and adopted the {\it Planck} 2018
    best-fit spatially-flat $\Lambda$CDM cosmology~\citep{2020A&A...641A...6P}.
    }
    \label{fig:p-Dtot-cutsky}
\end{figure}

In summary, the dispersions $\sigma_{\kin_x\kin_x}^2$,
$\sigma_{\kin_x\clu_x}^2$, $\sigma_{\clu_x\clu_x}^2$, etc, are given
in terms of linear combinations of the full-sky dispersions
$\sigma_{\kin\kin}^2$, $\sigma_{\kin\clu}^2$, and $\sigma_{\clu\clu}^2$,
(see Appendix~\ref{app:stats-cut} for full expressions), which in terms of the
angular power read
\begin{subequations}\label{eq:fullsky_dispersions}
\bea
    \sigkin^2 &= \frac{3}{4\pi}\ssp C\Zu{\ssp\kin\kin}_1 + \ckin^2\big({\sigma_\mrm{vir}}/{c}\big)^2, \\
    \sigint^2 &= \frac{3}{4\pi}\ssp \big(C\Zu{\ssp\clu\clu}_1 + {1}/{\bar{N}}\ssp\big), \\
    \sigkinint^2 &= \frac{3}{4\pi}\ssp C\Zu{\ssp\kin\clu}_1,
\eea
\end{subequations}
where $C\Zu{\ssp\kin\kin}_1$, $C\Zu{\ssp\clu\clu}_1$, and $C\Zu{\ssp\kin\clu}_1$ are evaluated
using equation~\eqref{eqn:Cab-ell} and the
appropriate transfer functions.
For the clustering dispersion we have also taken into account that number counts are
Poisson distributed, thus generating in $\sigint^2$ a
shot-noise contribution $C_1^\mrm{shot}=1/\bar{N}$,
where
$\bar{N}=N_\mrm{tot}/(4\pi f_\mrm{sky})$ is the
mean number-count density.
In this work ${N}_\mrm{tot}=1,355,352$ and here $f_\mrm{sky}=0.5$,
giving $\bar{N}=215,711$
(about 69 sources per square degree).
Note that with the kinematic transfer function~\eqref{eqn:Delta-kin-1}
the kinematic dispersion may also be written as
$\sigkin^2=\ckin^2\sp(\sigma\Z{R}^2+\sigma_\mrm{vir}^2)/c^2$, with
the usual (one-dimensional) velocity dispersion
\be
    \sigma\Z{R}\Zu2
    =\frac13\int^\infty_0\frac{k^2\dif k}{2\pi^2}\,\bigg(\frac{H_0 f_0}{k}\bigg)^2 W^2(kR)\sp P(k),
\ee
where $P(k)$ is the matter power spectrum.
The velocity dispersions considered in this work
are given in Table~\ref{tab:smooth}.
For each velocity dispersion we show in
Fig.~\ref{fig:p-Dtot-cutsky} the corresponding
PDF presented earlier [equation~\eqref{eqn:p-Dtot-cutsky}].
Note that in all PDFs the dispersions corresponding to shot noise and
clustering are fixed to the same values (they are independent of $R$).
However, the covariance between
clustering and kinematics (e.g.~$\sigma_{\kin_x\clu_x}^2$) varies
depending on $R$, as with the purely kinematic dispersion.

\begin{figure}
\centering
    \begin{subfigure}{\columnwidth}
        \centering
	    \includegraphics[width=\textwidth]{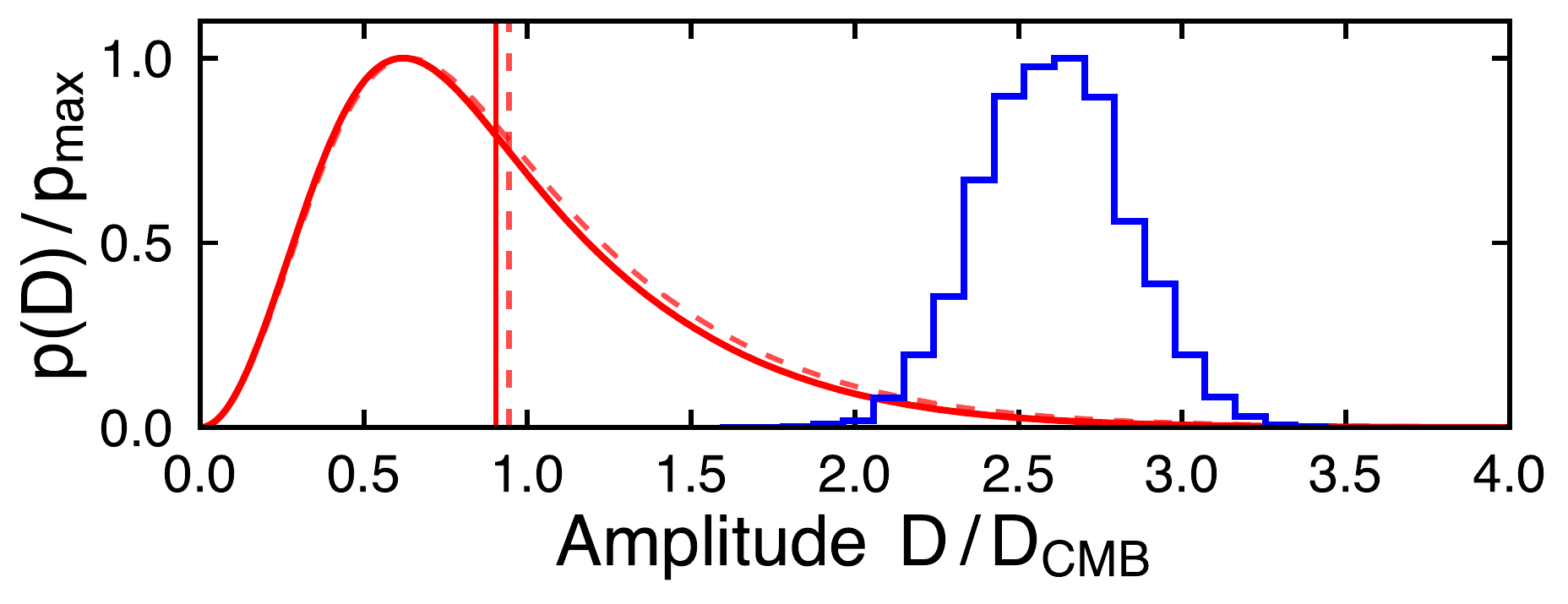}
    \end{subfigure}
    \caption{Comparison between the posterior and the marginalised theoretical
    prior~\eqref{eq:p-M_marg}. Here we use
    $M_\mrm{min}=10^{12}h^{-1}\ssp M_\odot$ and
    $M_\mrm{max}=10^{15}h^{-1}\ssp M_\odot$, and we include all
    contributions from clustering (which again have negligible impact).}
    \label{fig:pdf_mass_marg}
\end{figure}

Clearly our results will depend on the rather subjective value of $R$
(or $M$). Continuing our probabilistic approach, one way around this is
to simply marginalise over this uncertainty. In order to do this we need
to assign a prior $p(M)$, the distribution
of masses. This is given by the halo mass function $\dif n/\dif M$,
for which we use the mass function of \citet{1999MNRAS.308..119S}.
Thus the distribution of amplitudes, marginalised over all masses between
some interval, is
\be\label{eq:p-M_marg}
p(\Dtot)
=\int\dif M\ssp p(\Dtot\mid M) p(M)
=\int\dif M\ssp p(\Dtot\mid M) \frac{M}{\rho_m} \frac{\dif n}{\dif M}\, ,
\ee
where $p(\Dtot\mid M)$ is the probability distribution
given by equation~\eqref{eqn:p-Dtot-int} and
$\rho_m=\int\dif M\ssp M \dif n/\dif M$ (all matter in halos).
This distribution is shown in Fig.~\ref{fig:pdf_mass_marg}.
Compared with equation~\eqref{eqn:p-Dtot-cutsky}, this distribution
exhibits a slightly heavier tail; this is also reflected in
the difference between the mean and mode (cf.~Fig.~\ref{fig:p-Dtot-cutsky}).

\subsection{Effect of source evolution on $\dkin$}\label{sec:evol}
The kinematic dipole as given by equation~\eqref{eqn:d-kin} is idealised given that the source population likely evolves over time~\citep{2022MNRAS.512.3895D}.
Recall that this equation is based on a uniform sample of radio sources, each with identical spectral index $\alpha$, producing an integral
source count with constant slope $x$ at the flux density limit~\citep{1984MNRAS.206..377E}.
In practice, there will be some amount of population variance among the measured $\alpha$ (as found by S21 in their CatWISE
sample). Although this may be due to some intrinsic variation in AGN emission, it is also possible that it is in part due to $\alpha$ 
having some dependence on redshift.
Moreover, in a flux-limited survey, the magnification bias $s=2x/5$ generally
depends on redshift:
for a fixed flux threshold, the number of unobserved sources grows
as the luminosity threshold is increased (i.e.~the slope $x$ is an increasing function of redshift).

Revisiting the derivation of equation~\eqref{eqn:d-kin} in a
more general context, \citet{2022MNRAS.512.3895D}
showed that when allowing $x=x(z)$ and $\alpha=\alpha(z)$ the kinematic dipole becomes
\be\label{eqn:d-kin-intz}
\dkin=\bar{A}_\kin\sp\bm\beta,
\:\:\:\text{with}\:\:\:
\bar{A}_\kin\equiv\int\dif z\ssp p_z(z)\big[2+x(z)(1+\alpha(z))\big],
\ee
i.e.~the prefactor in the standard formula~\eqref{eqn:d-kin} is replaced by its average over the source redshift distribution.%
\footnote{Here we integrate along redshift instead of comoving distance, as done in \citet{2022MNRAS.512.3895D}; 
both expressions of $\dkin$, however, are equivalent at linear order.
Alternatively, instead of $\alpha(z)$, the integrated dipole can be 
expressed in terms of the evolution bias $f_\mrm{evo}(z)$, which parametrises how
a tracer's population number evolves over
time~\citep{2018JCAP...01..013M,2021JCAP...11..009N,2022MNRAS.512.3895D}.}
(Note that the velocity $\bm\beta$ is still given at the
observer's position.)
This integrated dipole generalises
the standard form, which is recovered when either
$x$ or $\alpha$ are redshift independent, or one observes at fixed redshift.

Evaluating the integrated dipole requires knowledge of
$x(z)$, which we here do not have without a measurement of the quasar
luminosity function~\citep{2020MNRAS.499.2598W,2022arXiv221204925G}.
Since $\bar{A}_\kin$ is given as the expectation
over the sample we will instead consider an equivalent but more suggestive form:
\be\label{eqn:d-kin-alt}
\bar{A}_\kin=\int\dif x\int\dif\alpha\ssp
p(x,\alpha)\big[2+x(1+\alpha)\big],
\ee
where $p(x,\alpha)$ is the joint distribution, for
which we have from S21 an empirical estimate of its marginal $p(\alpha)$ (based on  measurements of
$\alpha$ from {\it W1} and {\it W2} bands).
From this equation we have that [cf.~equation~\eqref{eqn:d-kin}]
\be\label{eqn:d-kin-corr}
\bar{A}_\kin
=\big[2+\bar{x}(1+\bar\alpha)\big]
    +\rho\sigma_x\sigma_\alpha
=\ckin+\rho\sigma_x\sigma_\alpha,
\ee
where $\bar{x}=\int\dif x\ssp xp(x)$ and
$\bar\alpha=\int\dif\alpha\ssp \alpha p(\alpha)$ denote the means;
$\sigma_x=\langle(x-\bar{x})^2\rangle^{1/2}$ and
$\sigma_\alpha=\langle(\alpha-\bar{\alpha})^2\rangle^{1/2}$ the variances;
and $\rho\equiv\langle(x-\bar{x})(\alpha-\bar{\alpha})\rangle/\sigma_x\sigma_\alpha$ the correlation coefficient between $x$ and $\alpha$.
Thus we have a correction to the standard formula
if $\langle x\alpha\rangle\neq\langle x\rangle\langle\alpha\rangle$, i.e.~if $\rho\neq0$.
Although we do not have $x(z)$ nor $p(x)$, we do know
its mean, $\bar{x}=1.7$, as obtained from the whole sample.
Based on this information alone we can nevertheless make
progress.
Continuing our Bayesian approach, we shall fix $p(x)$ according
to the \emph{principle of maximum entropy}~\citep{Jaynes_book},
that is, we choose for $p(x)$ that which maximises
the information entropy $\int\dif x\sp p(x)\ln p(x)$, subject
to the constraint of known mean $\bar{x}$
and the physical requirement that $x\geq0$. This leads us
to $p(x)=\lambda\rme^{-\lambda x}$, with $\lambda=1/\bar{x}$,
i.e.~the exponential distribution.
The mean is of course $\bar{x}$, and the variance
$\sigma_x^2=\bar{x}^2$.
Separately, using the empirical distribution of $\alpha$
obtained by S21, we determine $\bar\alpha=1.26$ and
$\sigma_\alpha=0.597$.

We can now estimate $\bar{A}_\kin$ using equation~\eqref{eqn:d-kin-corr}.
Though this requires the unknown correlation coefficient $\rho$,
we do however have an upper limit on it. Thus with $\rho=1$ we have
a maximum correction of $\rho\sigma_x\sigma_\alpha\leq1.7\times0.597=1.0$.
It should be noted that this is not a strict upper limit for it depends
on what one assigns to the rather uncertain $p(x)$.
With this caveat in mind, we estimate
$\bar{A}_\kin\leq6.7$, that is, we have a
correction to the standard prefactor ($\ckin=5.8$)
of no more than $17\%$. The distribution corresponding
to this upper limit is shown in Fig.~\ref{fig:p-Dtot-cutsky}
(see dashed curve), where we see that
the improvement in the tension is modest at best.
Clearly, in order to fully reconcile the observed
dipole amplitude in this manner we would
require a substantially larger correction.
But even with full knowledge of
both $x(z)$ and $\alpha(z)$, a correction of the needed size
seems unlikely, as mentioned in \citet{2022MNRAS.512.3895D}.

\subsection{Assessing the tension}\label{sec:numerical-results}
Fig.~\ref{fig:p-Dtot-cutsky} gives a side-by-side comparison of the
CatWISE posterior together with the theoretical probability
distribution of $\Dtot$ according to \LCDM\ (evaluated using the
empirical distributions of the redshifts and measured spectral indices).
The first point to note is that the clustering signal is much
weaker than
the kinematic signal. In particular, the contribution from
cross-correlations between the clustering and kinematic dipole
is roughly an order of magnitude larger than the clustering
signal (but still small compared to the kinematic signal).
Furthermore, we have set $\bqso=2$ using the
    \citet{2005MNRAS.356..415C} quasar bias parametrisation for our mean redshift
    $\bar{z}=1.2$, although we note that $p(\Dtot)$ is rather
    insensitive to the choice of $\bqso$ and time dependence, the main contribution coming from the kinematic terms.
Indeed, the fact that the clustering signal is small is not entirely
surprising: the quasar sample is made up  in large part of
sources mostly at moderate to high redshifts, with the vast majority ($99\%$) of sources
at $z>0.1$, so that a large clustering dipole from a local
population of sources is not expected.
Nevertheless, this confirms that, even with a more complete calculation of
the impact of clustering on the dipole (including magnification bias
among other effects), the kinematic contribution to the CatWISE dipole
remains the dominant component.

A comparison with the predictions on the full sky
(Appendix~\ref{app:Dtot-details}) shows that the masking
of the Galactic plane causes a significant loss of dipole power, and thus
a decrease in the expected dipole amplitude. 
This is evident if we compare $p(\Dtot)$ with the mask (Fig.~\ref{fig:p-Dtot-cutsky}) 
and without the mask (Fig.~\ref{fig:p-Dtot}): the presence of the mask
suppresses the tail probabilities, shifting the mean towards smaller values.
While the mixing of modes does result in a leakage of power from high multipoles into
the dipole, it is not nearly enough to compensate the reduction
in power of the dipole itself.
This is clear from the
shift of the distribution to smaller $\Dtot$, which makes
prominent a long tail by which rather large amplitudes can
still be accommodated by \LCDM.

\begin{figure}
\centering
    \begin{subfigure}{\columnwidth}
        \centering
	    \includegraphics[width=\textwidth]{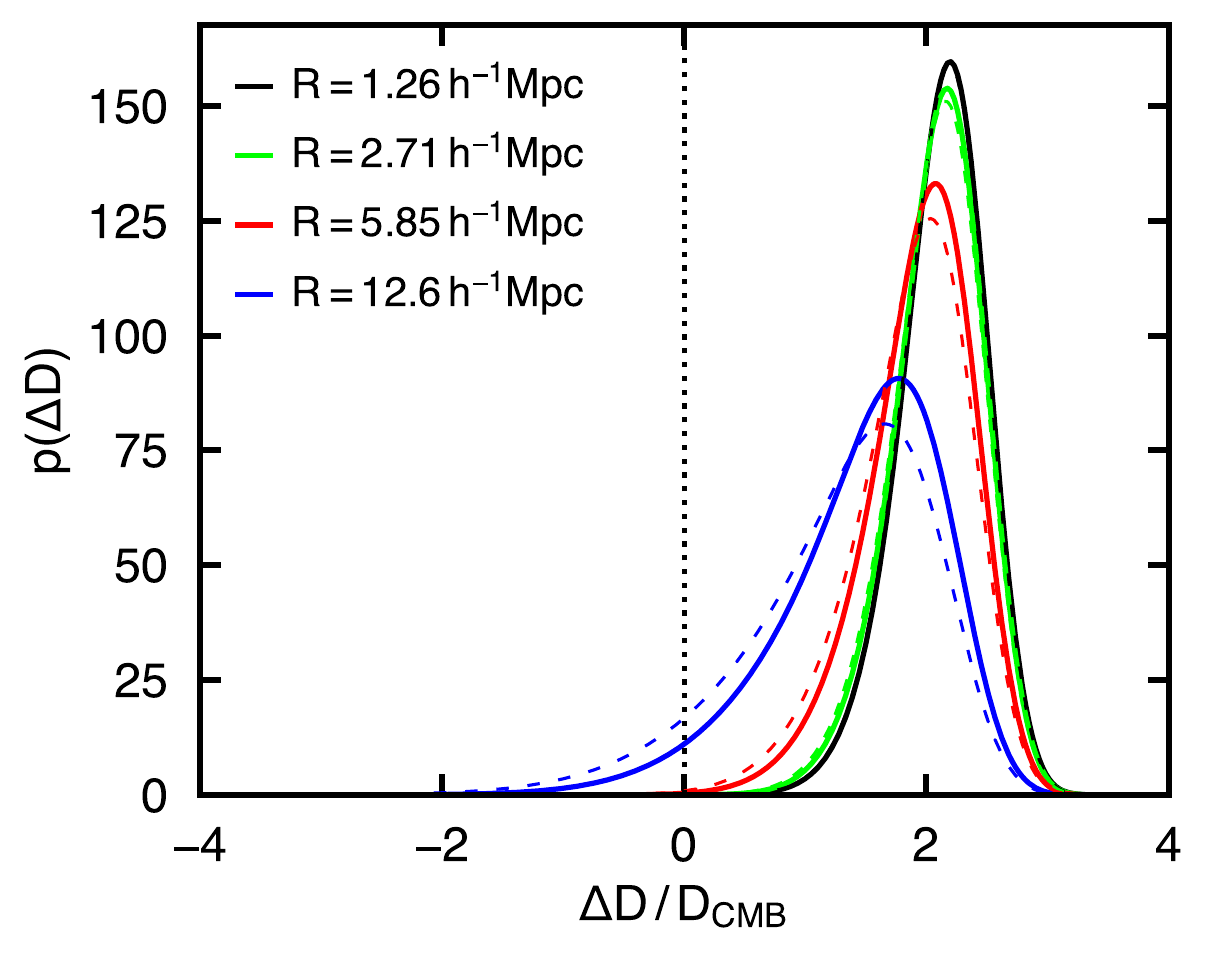} 
    \end{subfigure}
    \caption{Normalised shift distribution~\eqref{eqn:shift-pdf}
    derived from the
    \LCDM\ prior and posterior distributions for a range of smoothing lengths (as in Fig~\ref{fig:p-Dtot-cutsky}).
    Shown in dashed curves are the corresponding distributions with maximal
    source-evolution correction ($\rho=1$).}
    \label{fig:pdiff}
\end{figure}

To what extent is the \LCDM~prior compatible
with the posterior found? 
One way to summarise the discrepancy seen in Fig.~\ref{fig:p-Dtot-cutsky}
is to compute the `tension statistic'
$T\equiv|\mu_1-\mu_2|/(\sigma^2_{1}+\sigma^2_{2})^{1/2}$
for two distributions $p_1$ and $p_2$, with means $\mu_1$ and $\mu_2$,
and variances $\sigma_{1}^2$ and $\sigma_{2}^2$.
But this is a rather crude measure when at least one of the distributions
are non-Gaussian, as is the case for our prior~\eqref{eqn:p-Dtot-cutsky}.
A more comprehensive way to quantify the
discrepancy is to consider the PDF of the differences
$\Delta D=D_1-D_2$~\citep{2019PhRvD..99d3506R}:
\be\label{eqn:shift-pdf}
p(\Delta D)
=\int^\infty_{-\infty}\dif D\,p_1(D-\Delta D)p_2(D),
\ee
where $p_1(D_1)$ is the theoretical prior and $p_2(D_2)$ is the posterior.
Note that because there is zero probability of finding a negative amplitude, this is essentially a one-sided convolution.
(This integral is evaluated using a posterior smoothed by
kernel density estimation.) The result is shown in Fig.~\ref{fig:pdiff}.
Unsurprisingly, we see that $p(\Delta D)$ has
most support when $\Delta D$ is far from zero, peaking around
$\Delta D \simeq 2D_\mrm{CMB}$ (corresponding roughly
to the difference of the means of the prior and posterior
distribution). 
Thus the probability of finding any $\Delta D>0$ is given by
$P(\Delta D>0)=\int^\infty_0\dif\Delta D\ssp p(\Delta D)$.
From this probability we can compute a statistical significance
using the `rule of thumb' that the number of
sigma be equal to
$\sqrt2\,\mrm{erf}^{-1}(P(\Delta D>0))$.%
\footnote{This reduces to the aforementioned $T$ statistic in the case that
$p(\Delta D)$ is Gaussian.}
Depending on the choice of $R$, 
we find tensions at the level of
$5.8\sigma$, $5.5\sigma$, $4.3\sigma$, and
$2.1\sigma$ (from smallest to largest $R$).
Note that these tensions are only marginally reduced when including
the source-evolution corrections described in Section~\ref{sec:evol}.

By contrast when we marginalise over all masses between our lower and upper
limit, $M=10^{12}h^{-1}\ssp M_\odot$ and $M=10^{15}h^{-1}\ssp M_\odot$
(corresponding to $R=1.26h^{-1}\ssp\mrm{Mpc}$
and $R=12.6h^{-1}\ssp\mrm{Mpc}$), we find by equation~\eqref{eq:p-M_marg} a
tension of $3.8\sigma$ (and $3.6\sigma$ with source-evolution correction).
This is still a considerable tension but less severe than suggested above. The
slight easing of the tension is due to the fact that the tail probabilities
of this distribution are larger than that of the corresponding distribution
without marginalisation.
This is evident in Fig.~\ref{fig:pdf_mass_marg}, which shows
a small overlap of the tail with the posterior.

\section{Conclusions}\label{sec:conclusions}
We have reanalysed the CatWISE quasar sample of \citet[][S21]{2021ApJ...908L..51S} using
a Bayesian approach.
By comparison of several hypotheses, we found that the data was best described
by a  dipole with amplitude $\Dtot=(19\pm2)\times10^{-3}$ and direction
$(l,b) = ( 237\dotdeg2^{+7\dotdeg9}_{-8\dotdeg0} , 41\dotdeg8\pm5\dotdeg0)$.
Several parametric models were investigated.
With the dipole amplitude free to vary, this dipole direction is only mildly preferred over
a dipole aligned with the CMB, more so than that found by S21.
However, it is important to note that
the \emph{amplitude} is considerably larger than expected from the conventional kinematic interpretation:
accepting that the CMB dipole is produced entirely by our kinematic motion, our dipole
converts to a speed of $1002\,\mrm{km\,s^{-1}}$ relative to
the CMB frame, a factor of $2.7$ 
larger than the putative speed of $370\,\mrm{km\,s^{-1}}$.
Taking into account the full posterior distribution (marginalising over all other parameters) we find a discrepancy at the $5.7\sigma$ level.
While our inferred amplitude is somewhat larger
than that found by S21 using a different
estimation technique ($\Dtot=15.54\times10^{-3}$),
the qualitative conclusion of an anomalously large amplitude (with
direction in good agreement with that of the CMB dipole) appears to
be robust to methodological differences of analysis.

A comparison of the posterior with the theoretical prior indicates
the extent of the discrepancy for \LCDM.
In order to perform this comparison we
applied a mask to the underlying number-count fluctuations,
imposing the same Galactic plane cut as in the analysis. 
The mask was found to have a considerable effect on the
likely values of the dipole amplitude and its distribution.
In particular, we
found that the full-sky distribution
overestimates the likely values of the amplitude, worsening
the tension.
Additionally, we have considered a redshift dependence in the spectral
properties of the kinematic dipole, finding a slight easing
of the tension -- though we cannot claim with confidence that such an
effect is small, given the limitations of our analysis relating to
the unknown CatWISE distribution (or redshift dependence) of the
magnification bias.
But based on our present findings we conclude that the CatWISE
dipole remains a puzzle for the standard model of cosmology.

\section*{Note Added}
Since this manuscript was submitted, a work by \citet*{2023A&A...675A..72W} appeared in which
the radio dipole is estimated from the Rapid ASKAP Continuum Survey (RACS) and the
NRAO VLA Sky Survey (NVSS) catalogue data, extending the Bayesian method
presented here to take into account systematics relevant to these catalogues.
From a combined analysis of RACS and NVSS, they find a dipole with direction
perfectly aligned with the CMB but with amplitude exceeding the expected value
by a factor of about three at $4.8\sigma$, in good agreement with the
results reported here.

\section*{Acknowledgements}
We thank \citet{2021ApJ...908L..51S}
for making public their data and analysis code, and the
anonymous referee for a helpful report.
LD thanks Camille Bonvin for useful discussions on
the effect of source evolution.
LD was supported by the Australian Government Research Training Program.
A preliminary version of this work appeared in
\cite{2123-27817}.
GFL received no financial support for this research.
This research made use of the Python packages
{\sc Numpy}~\citep{numpy},
{\sc Matplotlib}~\citep{matplotlib},
\healpix~\citep{2005ApJ...622..759G,Zonca2019},
{\sc corner}~\citep{corner},
and \mbox{\sc GetDist}~\citep{GetDist}.
Initial explorations of the posterior probability space were made using
{\sc emcee}~\citep{ForemanMackey:2012ig} and
{\sc dynesty}~\citep{2004AIPC..735..395S,10.1214/06-BA127,cite-key,2020MNRAS.493.3132S}.

\section*{Statement of Contribution}
The statistical analysis of the CatWISE quasar sample, including the exploration
of the posterior distributions and calculation of the Bayesian evidence was primarily 
undertaken by GFL and BJB, after preliminary analysis with LD. The calculation
of the dipole's likely amplitude and the comparison with the
posterior distributions was undertaken by LD. All authors contributed to the interpretation of the results and 
the writing of the paper.

\section*{Data Availability}
The data and code used in this project will be made available upon reasonable request to the authors. 



\bibliographystyle{mnras}
\bibliography{paper} 

\appendix
\section{Details on the theoretical dipole}\label{app:b-1M-derivation}
Here we compute the kinematic and clustering dipoles
in terms of the harmonic transfer functions,
$\Delta\Zu{\kin}_\ell$ and $\Delta\Zu{\clu}_\ell$, respectively.
First, we recall that for a general field $O(\n)$ the coefficients
$a_{\ell m}$ can always be expressed as a $\bk$-space integral over the primordial perturbations,
weighted by the appropriate harmonic transfer function $\Delta_\ell(k)$:
\be
    a\Z{\ell m}
        =(-\im)^\ell\sp4\pi\int\frac{\dif^3\bk}{(2\pi)^3}\:
            Y^*_{\ell m}(\hat\bk)\sp \Delta_\ell(k)\sp {\zeta}(\bk).
    \label{eqn:a-1M}
    \ee
Then given the transfer functions for any two observables, labelled $A$ and $B$,
the spectrum $C_\ell^{AB}$ is given by
\be
    C_\ell^{AB}
        = 4\pi\int^\infty_0\frac{\dif k}{k}\:
        	\Delta\Zu{A}_\ell(k)\sp \Delta\Zu{B}_\ell(k)\sp\PR(k),
    \label{eqn:C-ell-ab}
\ee
where we assume a standard primordial (adiabatic) power spectrum $\PR(k)$ defined by
$\langle{\zeta}(\bk){\zeta}^*(\bk')\rangle=(2\pi)^3\delta_{\rm D}(\bk-\bk')\sp 2\pi^2\PR(k)/k^3$. In principle, superhorizon isocurvature fluctuations can also be considered, which are known to give rise to an
intrinsic dipole in the CMB~\citep{Turner:1991dn}. However, it was
recently shown~\citep{Domenech:2022mvt} that such modes cannot also
induce a sizeable intrinsic dipole in the number counts.

Now, using equations~\eqref{eqn:a-1M} and \eqref{eqn:dipole-vecs}, we find that the model
prediction for the dipoles $\mbf{d}_\mrm{X}$, $\mrm{X}\in\{\kin,\clu\}$, can be written as
\be
    \mbf{d}_\mrm{X}
        = -\ssp4\pi\im\sqrt{\frac{3}{4\pi}}
          \int\frac{\dif^3\bk}{(2\pi)^3}\, \mat{U}^\dagger\ssp\bm{Y}_1(\hat\bk)\ssp\Delta\Zu{\mrm{X}}_1(k)\ssp {\zeta}(\bk),
\ee
where $\bm{Y}_1\equiv(Y_{1-1},Y_{11},Y_{10})\Zu\T$ and
we have the unitary matrix
\be\label{eq:U}
    \mathbf{U}
    =
    \begin{pmatrix}
        \!\phantom{-}1/\sqrt{\ssp2}    &   \im/\sqrt{\ssp2}  &   0   \\[2pt]
        \!          -1/\sqrt{\ssp2}    &   \im/\sqrt{\ssp2}  &   0   \\[2pt]
        \!0           &       0        &   1
    \end{pmatrix},
\ee
with $\mat{U}\mat{U}^\dagger=\mat{U}^\dagger\mat{U}=\mat{I}_3$.
Note that $\mat{U}^\dagger\ssp\bm{Y}_1$ is a vector containing
the $\ell=1$ real-valued spherical harmonics.

Computing the statistics of
$\mbf{d}_\mrm{X}$ -- i.e.~$\langle{d}_{\mrm{X}i}{d}_{\mrm{Y}j}\rangle=\sigma^2\Z{\mrm{XY}}\sp \delta_{ij}^\mrm{K}$ -- is now straightforward.
To obtain the dispersion $\sigma^2\Z{\mrm{XY}}$ we can
simply select the $z$-component, all components being statistically equivalent
under isotropy. Using orthogonality of spherical harmonics,
we have
\bea
    \sigma\Z{\mrm{XY}}^2
        &=3\int^\infty_0\frac{\dif k}{k}\ \Delta_1\Zu{\mrm{X}}(k)\ssp \Delta_1\Zu{\mrm{Y}}(k)\ssp \PR(k)
        =\frac{3}{4\pi}\ssp C\Zu{\mrm{XY}}\Z1,
    \label{eqn:sig-ab}
\eea
where we have used
$\langle{\zeta}(\bk)\ssp{\zeta}^*(\bk')\rangle=(2\pi)^3\delta_{\rm D}(\bk-\bk')\sp 2\pi^2\PR(k)/k^3$.

Note that in this appendix the $a_{\ell m}$'s belong to the full sky.
Those belonging to the cut sky can always be obtained
in terms of the full-sky $a_{\ell m}$'s by geometric 
linear relations.
The explicit expressions are given later
in Appendix~\ref{app:mode-coupling}.

\subsection{Clustering transfer function} 
Clustering gives rise to fluctuations $\delta_\qso(\n)$ in the number counts. In spherical harmonics,
\be
	\delta_\qso(\n)
	=\sum_{\ell,m}^\infty a\Z{\ell m}\, Y_{\ell m}(\n),
\quad
	a\Z{\ell m}=\int\dif^2\n~Y_{\ell m}^*(\n)\ssp\delta_\qso(\n).
\ee
Here we are interested in the dipole $\dint\cdot\n=\sum_{m=-1}^1 a_{1m}\sp Y_{1m}(\n)$.
Substituting $\delta_\qso(\n)=\int\dif z\,p_z(z)\ssp\delta_\qso(\n,z)$, with $p_z$ the
normalised redshift distribution, and the Fourier
representation of $\delta_\qso(\n,z)$, we have
\begin{align*}
a\Z{\ell m}
&=\int\dif^2\n~Y_{\ell m}^*(\n)\int^\infty_0\dif z~p_z(z)\ssp\delta_\qso(\n,z) \nonumber\\[-2pt]
&=\int\dif^2\n~Y_{\ell m}^*(\n)\int^\infty_0\dif z~p_z(z)\int\frac{\dif^3\bk}{(2\pi)^3}~
	{\delta_\qso}(\bk,z)\ssp \rme\Zu{-\im\bk\cdot\chi(z)\n},
\end{align*}
where in the second line we write $\x=\chi(z)\n$.
Now substituting the plane-wave expansion,
$\rme^{-\im\bk\cdot\chi\n}
=\sum_{\ell,m}(-\im)^{\ell}4\pi\sp j_{\ell}(k\chi)\sp
Y^*_{\ell m}(\hat\bk)\sp Y^{\phantom*}\Z{\ell m}(\n)$,
into the foregoing expression we find
\begin{align*}
a\Z{\ell m}
&=\int\dif^2\n~Y_{\ell m}^*(\n)
	\int^\infty_0\dif z~p_z(z)\int\frac{\dif^3\bk}{(2\pi)^3}~{\delta_\qso}(\bk,z) \nonumber\\
	&\qquad\qquad\qquad\times\sum_{\ell',m'}(-\im)^{\ell'}4\pi\ssp j_{\ell'}\big(k\chi(z)\big)\ssp
	Y^*_{\ell'm'}(\hat{\bk})\ssp Y_{\ell'm'}(\n) \nonumber\\
&=(-\im)^\ell\sp 4\pi\int\frac{\dif^3\bk}{(2\pi)^3}\ Y^*_{\ell m}(\hat{\bk}) \nonumber\\
	&\qquad\:\times\bigg[\int^\infty_0\dif z~p_z(z)\ssp \bqso(z)\ssp\mathcal{T}_\delta(k,z)\ssp
			j_\ell\big(k\chi(z)\big)+\cdots\bigg]\zeta(\bk),
\end{align*}
where in the second line we used the
orthogonality of spherical harmonics.
Here we have inserted
$\delta_\qso(\bk,z)=\bqso(z)\ssp\delta(\bk,z)+\cdots=\bqso(z)\ssp\mathcal{T}_\delta(k,z)\ssp\zeta(\bk)+\cdots$,
where the ellipsis are corrections from 
redshift-space distortions, Doppler effects, and general-relativistic
corrections 
\citep[see, e.g.,][appendix A, for complete expressions]{CLASSgal}.
We thus have
\be
	\Delta\Zu{\clu}_\ell(k)
	=\int^\infty_0\dif z~p_z(z)\ssp\bqso(z)\ssp\mathcal{T}_\delta(k,z)\ssp
			j_\ell\big(k\chi(z)\big)
   +\cdots.
\ee

\subsection{Kinematic transfer function}\label{app:calc-b1m}
For the kinematic dipole recall that $\dkin\cdot\n=\ckin\,\bm\beta\cdot\n$, with
$\ckin=2+x(1+\alpha)$ and $\bm\beta=\v_O/c=\v(\x\!=\!0)/c$. In harmonics,
\be
	\dkin\cdot\n
	=\sum_{L,M} b\Z{LM}\, Y_{LM}(\n),
\quad
	b\Z{LM}=\int\dif^2\n~Y_{LM}^*(\n)\ssp\dkin\cdot\n,
\ee
the calculation carries through in much the same way as with the previous calculation for number
counts, {provided} that we keep $\x$ arbitrary, setting $\x=0$ only at the end of the
calculation. (Note that since we are dealing with a pure dipole only
$L=1$ coefficients are nonzero.) We thus write equation~\eqref{eqn:b-LM} as
\bea
    b_{1M}
        &=\int\dif^2\n\ Y_{1M}^*(\n)\ssp\ckin\,\v\Z{R}(\x\!=\!0)\cdot\n/c \nonumber\\[-2pt]
        &=\int\dif^2\n\ Y_{1M}^*(\n)\ssp\ckin
            \int\frac{\dif^3\bk}{(2\pi)^3}
            \frac{-\im\bk\cdot\n}{k}\bigg(\frac{H_0 f_0}{c\sp k}\bigg)\sp W(kR)\sp\delta(\bk)
    \label{eqn:b-1M-w1}
\eea
and then use that with $\x=\chi\n$, where $\chi$ the comoving distance, the dipole 
$-\im\sp\bk\cdot\n$ can be expressed as 
$-\im\sp\bk\cdot\n=(\partial\sp\rme^{-\im\bk\cdot\chi\n}/\partial\chi)|_{\chi=0}$.
Now substituting the plane-wave expansion
into equation~\eqref{eqn:b-1M-w1} we find
\begin{align*}
    b_{1M}
        &=\int\dif^2\n\ Y_{1M}^*(\n)\sp\ckin
            \int\frac{\dif^3\bk}{(2\pi)^3}
            \bigg(\frac{H_0f_0}{c\sp k}\bigg) W(kR)\sp\delta(\bk) \nonumber\\
        &\qquad\qquad\times
            \frac1k\frac{\partial}{\partial\chi}
            \bigg[\sum_{\ell,m}(-\im)^{\ell}4\pi\sp j_\ell(k\chi)\sp
                    Y^*_{\ell m}(\hat\bk)\sp Y\Z{\ell m}(\n)\bigg]_{\chi=0} \nonumber\\[2pt]
        &=(-\im)\sp4\pi\int\frac{\dif^3\bk}{(2\pi)^3}\:
            Y^*_{1M}(\hat\bk)\sp \ckin\bigg(\frac{H_0f_0}{c\sp k}\bigg)W(kR)\sp\delta(\bk)
                \frac1k\frac{\partial j_1}{\partial\chi}\bigg|_{\chi=0},
\end{align*}
where in the second line we have used the orthogonality of spherical
harmonics.
Since $j_1(x)=x/3+\mathcal{O}(x^3)$, the derivative with respect to $\chi$, at $\chi=0$,
exactly evaluates to $k/3$; inserting $\delta(\bk)=\mathcal{T}_\delta(k,z)\sp{\zeta}(\bk)$, we finally get
\be\nonumber
    b_{1M}
        =(-\im)\sp4\pi\int\frac{\dif^3\bk}{(2\pi)^3}\:
            Y_{1M}^*(\hat\bk)
                \bigg[\frac13\sp\ckin\sp W(kR)\bigg(\frac{H_0f_0}{c\sp k}\bigg)\sp \mathcal{T}_\delta(k)\bigg]\sp{\zeta}(\bk),
\ee
where $\mathcal{T}_\delta(k)$ is evaluated at $z=0$. Note that
$\bqso$ does not appear here because $\v$ is sourced by
matter perturbations. Since this expression is in the form of equation~\eqref{eqn:a-1M} we can immediately
read off the kinematic transfer function from
the contents of the square brackets:
\be
    \Delta\Zu{\kin}_1(k)
        =\frac13\sp\ckin\sp W(kR)\sp\bigg(\frac{H_0f_0}{c\sp k}\bigg)\sp\mathcal{T}_\delta(k).
    \label{eqn:Delta-kin-1}
\ee
Note that because we evaluate $\v$ at $\x=0$ the 
transfer function $\mathcal{T}_\delta(k)$ is simply evaluated
at $z=0$ and not integrated. Further note that to
include source evolution,
simply replace $\ckin$ with its average $\bar{A}_\kin$ 
given by equation~\eqref{eqn:d-kin-corr}.

\section{Pixelization}
We can also consider the effect of the pixelization on the theoretical power.
At the field level, the pixelization of the number-count fluctuations is given by
\be
    \delta\Z{\qso,i} = \int\dif^2\n\: W^\mrm{pix}_i(\n)\ssp \dqso(\n),
\ee
where $W^\mrm{pix}_i(\n)$ is the \healpix\ pixel window function for the $i$th pixel,
which is equal to zero, unless $\n$ falls within the $i$th pixel, in which case
$W^\mrm{pix}_i$ is equal to $1/\Omega_{\rm pix}$, with $\Omega_{\rm pix}$ being the
pixel area. This window function is normalised,$\int\dif^2\n\,W^\mrm{pix}_i(\n)=1$,
so that $\delta\Z{\qso,i}$ is the average fluctuation in the $i$th pixel.
Note that because the \healpix\ pixel shape varies azimuthally there is no global pixel window function
that applies to all pixels. Provided a large enough $N_\mrm{side}$ is chosen this is not really a problem: the differences between pixel window functions
only becomes important when considering large $\ell$, but this can always be remedied by choosing
a larger $N_\mrm{side}$.

We will ignore the azimuthal variation in pixels,
as is routinely done. Under this
approximation pixelized fields are also statistically isotropic and the angular power spectrum
of the pixelized field is
$
    C^\mrm{\ssp pix}_\ell = (\sp\bar{W}^\mrm{\ssp pix}_\ell\sp)^2 C_\ell
$,
where $C_\ell$ is the unpixelized power, and $\bar{W}^\mrm{\sp pix}_\ell$ is the \healpix
pixel-averaged window function~\citep{2005ApJ...622..759G}. Since $\bar{W}^\mrm{\sp pix}_\ell\leq1$ there is a loss of power for all
$\ell$ (more for larger $\ell$).
In this work $N_\mrm{side}=64$ with which we find that
$\bar{W}^\mrm{\ssp pix}_\ell$ is equal to unity to
within $0.5\%$ for $\ell=0$ to $\ell=20$. As this power loss is negligible we have thus
ignored the effects of pixelization on the theoretical power.


\section{Derivation of $\Pr(\Dtot)$}\label{app:pD_derivation}
In this appendix we derive the probability distribution function $p(\Dtot)$ in both the full-sky and cut-sky regimes. The PDF in the former case is well known; it is the Maxwellian. In the latter case, considering an arbitrary symmetric Galactic
plane cut (the details of which will be fixed in Appendix~\ref{app:gal-cut}),
we derive another long-tailed PDF, one that is suppressed relative to the Maxwellian.
The calculation comes down to the marginalisation~\eqref{eqn:p-Dtot-int}, which we evaluate through the moment generating function. The full-sky PDF is parametrised by $\sigkin$,
$\sigint$, and $\sigkinint$; whilst the cut-sky PDF is parametrised by
$\sigma_{\kin_x\kin_x}$, $\sigma_{\kin_z\kin_z}$,
$\sigma_{\clu_x\clu_x}$, $\sigma_{\clu_z\clu_z}$, $\sigma_{\kin_x\clu_x}$, and
$\sigma_{\kin_z\clu_z}$ (these are, however, linearly related to those of the full sky).
The calculation is based on geometrical considerations;
no cosmological model needs to be assumed.

\subsection{Full sky}\label{app:Dtot-details}
To set up the problem, construct the six-dimensional vector $\bm{X}=(\dkin,\dint)\Zu\T$,
with $\bm{X}\sim\calN(\mbf{0},\cov)$.
By statistical isotropy only components along the same axis can
be correlated ($x$-$x$, $y$-$y$, etc).
Thus $\cov$ can be written in block matrix form~\eqref{eqn:cov-6x6}, with each block 
proportional to the $3\times3$ identity matrix,
$\langle\mathbf{d}_\mrm{X}\Zu{\phantom\T} \mathbf{d}_\mrm{Y}\Zu\T\rangle\propto\mathbf{I}_3$,
with $\mrm{X},\mrm{Y}\in\{k,c\}$.
In particular, we have that the expected amplitude of the kinematic dipole is
$\Dkin\equiv\langle\dkin^2\rangle^{1/2}=\sqrt{\sp3}\ssp\sigkin$,
and the expected amplitude of the clustering dipole is
$\Dint\equiv\langle\dint^2\rangle^{1/2}=\sqrt{\sp3}\ssp\sigint$,
where the root three is due to $\sigkin$ and $\sigint$ being
one-dimensional dispersions, while $\Dkin$ and $\Dint$ is the total dispersion. Evaluating $\sigkin$, $\sigint$,
and $\sigkinint$ requires specifying a cosmological model, but these are simply related
to the $\ell=1$ multipole of certain angular power spectra by $\sigma\Z{\mrm{XY}}\Zu2=3C^\mrm{XY}_1/(4\pi)$.

Though $\cov$ is a $6\times6$ matrix there are only three parameters needed to specify it (on
account of isotropy). In analysing this matrix we find it convenient to therefore
write it as a Kronecker product:
\be
    \cov
        =
        \begin{pmatrix}
            \,\sigkin^2 & \sigkinint^2\, \\[3pt]
            \,\sigkinint^2 & \sigint^2\,
        \end{pmatrix}
        \otimes \mathbf{I}_3,
    \qquad
    \cov^{-1}
        =
        \begin{pmatrix}
            \,\sigkin^2 & \sigkinint^2\, \\[3pt]
            \,\sigkinint^2 & \sigint^2\,
        \end{pmatrix}^{-1}
        \otimes \mathbf{I}_3,
\ee
where $\mathbf{I}_3$ is the $3\times3$ identity matrix, and the inverse $\cov^{-1}$ is
the Kronecker product of the inverse of the two matrices.%
\footnote{$\mat{A}^{-1}=(\mat{B}\otimes\mat{C})^{-1}=\mat{B}^{-1}\otimes\mat{C}^{-1}$.}
Thus we see that the covariances in $\cov$ are essentially contained in a $2\times2$ matrix.

The square of the amplitude of the total dipole, $Y\equiv\Dtot^2=|\dkin+\dint|^2$, can be
recast as the quadratic form
\be
    Y=\bm{X}\Zu\T\mathbf{A}\bm{X},
    \qquad
    \mathbf{A}
    \equiv
        \begin{pmatrix}
            \,\mathbf{I}_3 & \mathbf{I}_3\, \\
            \,\mathbf{I}_3 & \mathbf{I}_3\,
        \end{pmatrix}
        =
        \begin{pmatrix}
            \sp1 & 1\, \\
            \sp1 & 1\,
        \end{pmatrix}
        \otimes\mathbf{I}_3,
\ee
where $\mathbf{A}$ is a $6\times6$ symmetric matrix.
The moment generating function
${M}\Z{Y}(t)\equiv\langle \rme^{tY}\rangle$ can be written using the
`law of the unconscious statistician' as
\be
    {M}\Z{Y}(t)
        \equiv\int\dif Y\, \rme^{tY} p(Y)
        =\int\dif^6\bm{X}\, \rme^{t\bm{X}\Zu\T\mathbf{A}\bm{X}} p(\bm{X}),
    \label{eqn:M-Y}
\ee
i.e.~in terms of the probability distribution of $\bm{X}$.
Before we compute the second integral, we recall that ${M}_Y(-t)$ is the
(unilateral) Laplace transform
$\mathscr{L}$ of the PDF $p(Y)$:
\be
        \mathscr{L}[p(Y)](t)
        \equiv\int^\infty_0 \dif Y\, \rme^{-tY} p(Y)
        ={M}\Z{Y}(-t).
\ee
Therefore, once the moment generating function is in hand, and the sign of its argument
is flipped, we recover the PDF by simply taking the inverse Laplace transform:
\be
    p(Y)
        =\mathscr{L}^{-1}\big[{M}\Z{Y}(-t)\big](Y)
        \equiv\frac{1}{2\pi\im}\int^{\gamma+\im\infty}_{\gamma-\im\infty}\dif t\,
            \rme^{tY}{M}\Z{Y}(-t).
\ee
(Here $\gamma$ is some large, positive constant, chosen
so that the poles lie to the
left of it in the complex plane.)
In practice, we will not have to perform such integrals as we can build the solution
out of simpler, known transforms, by using the convolution properties of the Laplace transform.%
\footnote{Perhaps the more obvious approach to obtain the PDF is to use the characteristic function
$\phi\Z{Y}(t)\equiv{M}\Z{Y}(\im t)$ together with the Fourier transform.
However, this proves slightly awkward, since $Y\geq0$ forces us to consider a
one-sided Fourier transform.}

To evaluate the second integral in equation~\eqref{eqn:M-Y}, we note that it has the basic form
of a Gaussian convolution; because one `Gaussian' is without its normalisation factor, 
we find after some straightforward matrix algebra
\be
    {M}\Z{Y}(t)
        =\det\big(\mathbf{I}_6 - 2\sp t\mathbf{A}\cov\big)^{-1/2}
        =\prod_{j=1}^6 \lambdaup\Z{j}^{-1/2},
\ee
where in the second equality we have expressed the determinant in terms of the
eigenvalues $\lambdaup_j$ of $\mathbf{I}_6-2\sp t\mathbf{A}\cov$. Now specialising to
the case of statistical isotropy, we find that there are only two
distinct eigenvalues, which we call $\lambdaup_1$ and $\lambdaup_2$.
To see this we exploit the properties of the Kronecker product: write
$\mathbf{I}_6=\mathbf{I}_2\otimes\mathbf{I}_3$ and
$\mathbf{A}\cov=\bm\Sigma\otimes\mathbf{I}_3$, with
\be
    \bm\Sigma
        \equiv
        \begin{pmatrix}
            \sp1 & 1\, \\
            \sp1 & 1\,
        \end{pmatrix}
        \begin{pmatrix}
            \sp\sigkin^2 & \sigkinint^2\, \\[3pt]
            \sp\sigkinint^2 & \sigint^2\,
        \end{pmatrix}
        =
        \begin{pmatrix}
            \,\sigkin^2+\sigkinint^2 & \sigint^2+\sigkinint^2\, \\[4pt]
            \,\sigkin^2+\sigkinint^2 & \sigint^2+\sigkinint^2\,
        \end{pmatrix}.
\ee
Because
$\mathbf{I}_6-2\sp t\mathbf{A}\cov
    =(\mathbf{I}_2-2\sp t\bm\Sigma)\otimes\mathbf{I}_3$, there are two distinct eigenvalues
(each having multiplicity three) corresponding to the $2\times2$ matrix $\mathbf{I}_2-2\sp t\bm\Sigma$.
The eigenvalues are $\lambdaup_1=1$ and $\lambdaup_2=1-2\sp t\Delta^2$,
with
\be
    \Delta^2
        \equiv\sigkin^2 + 2\sp\rho\sp\sigkin\sp\sigint + \sigint^2.
\ee
Therefore, we have simply that ${M}_Y(t)=(1-2\sp t\Delta^2)^{-3/2}$, which
can be seen to closely resemble the moment generating function of the generalised
chi-squared distribution.
As mentioned, the PDF is recovered by inverse Laplace transform of ${M}_Y(-t)$:%
\footnote{The integral can be done with help of the convolution theorem, which
says that a product of moment generating functions corresponds to the convolution of their
individual PDFs.
Thus we write $(1-2\sp t\Delta^2\big)^{-3/2}=\prod_{j=1}^3(1-2\sp t\Delta^2\big)^{-1/2}$,
and note that the chi-squared distribution with one degree of freedom has a moment
generating function $(1-2\sp t\Delta^2\big)^{-1/2}$. Then the PDF
corresponding to ${M}_Y(t)$ is the convolution $p(Y)=p_x\ast p_y\ast p_z$ of
three such chi-squared distributions, which yields another chi-squared distribution
but with three degrees of freedom.}
\be
    p(Y)
        =\frac{1}{2\pi\im}\int^{\gamma+\im\infty}_{\gamma-\im\infty}
            \frac{\rme^{tY}\dif t}{\big(1+2\sp t\Delta^2\big)^{3/2}}
        = \frac{\sqrt{\sp Y}\sp \rme^{-Y/(2\Delta^2)}}{\sqrt{\sp2\pi}\ssp\Delta^3}.
\ee
Finally, the distribution for the amplitude $\Dtot\equiv\sqrt{\sp Y}$ is
${p(\Dtot)=p(Y\!=\!\Dtot^2)\ssp|\sp\dif Y/\dif\Dtot|}$, or, explicitly, after some algebra,
\be
    p(\Dtot)\ssp\dif\Dtot
        =\frac{4}{\sqrt\pi}\ssp\exp\bigg(-\frac{\Dtot^2}{2\Delta^2}\bigg)\ssp
        \frac{\Dtot\Zu2}{2\sp\Delta\Zu2}\sp\frac{\dif \Dtot}{\sqrt2\sp\sp\Delta}.
    \label{eqn:p-X}
\ee
This PDF, a chi distribution with three degrees of freedom (or Maxwellian), 
has mean $\langle\Dtot\rangle=\sqrt{\sp8/\pi\sp}\sp\Delta$ and
variance $\langle\Dtot^2\rangle=3\Delta^2$. However, the variance
is larger than might be expected from a standard Maxwellian due to cross terms
($\sigma^2_{\kin_x\clu_x}$, $\sigma^2_{\kin_z\clu_z}$, etc).

\begin{figure}
\centering
    \begin{subfigure}{\columnwidth}
        \centering
	    \includegraphics[width=\textwidth]{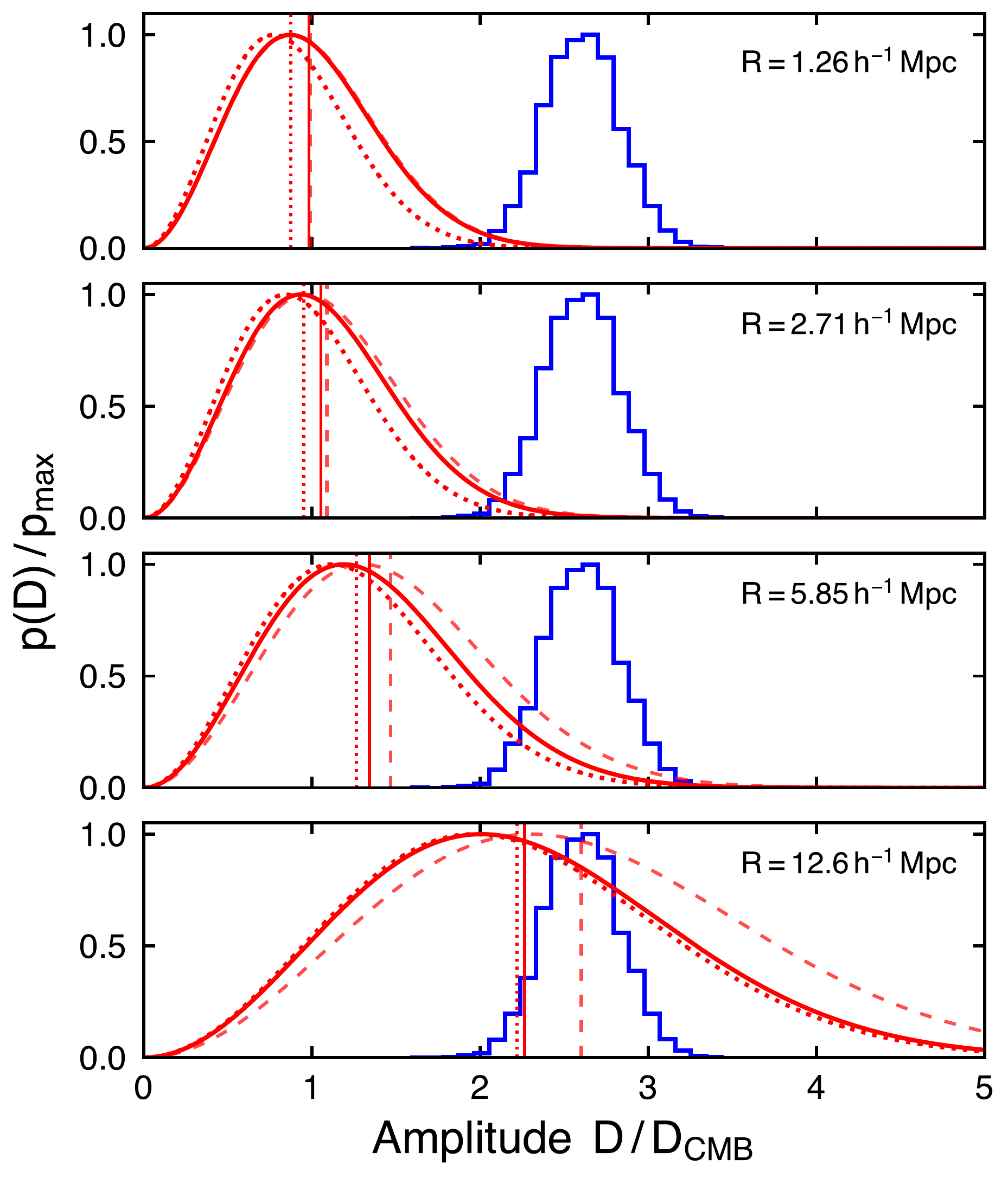} 
    \end{subfigure}
    \caption{As for Fig.~\ref{fig:p-Dtot-cutsky} but on the full sky ($f_\mrm{sky}=1$).
    The probability distribution is simply the Maxwellian~\eqref{eqn:p-X}
    with dispersions
    given by equations~\eqref{eq:fullsky_dispersions}. The main point here is
    that without the mask larger values of $\Dtot$ are more probable compared
    with the case of including the mask}.
    \label{fig:p-Dtot}
\end{figure}

\subsection{Cut sky}\label{app:p-Dtot-cutsky}
The Galactic plane cut, described later in Appendix~\ref{app:galactic-cut}, has the effect of identifying a preferred direction, i.e.~along the $z$-axis (where the $z$-axis is orthogonal to the Galactic plane).
Because of this the covariance now
takes the form
\bea\label{eqn:cov-6x6-cutsky}
    \cov
    &=
        \left(
        \begin{array}{cccccc}
            \sigma^2_{\kin_x\kin_x} & 0 & 0 & \,\sigma^2_{\kin_x\clu_x} & 0 & 0 \\
            0 & \sigma^2_{\kin_x\kin_x} & 0  & \,0 & \sigma^2_{\kin_x\clu_x} & 0 \\
            0 & 0 & \sigma^2_{\kin_z\kin_z} & \,0 & 0 & \sigma^2_{\kin_z\clu_z} \\
            \sigma^2_{\kin_x\clu_x} & 0 & 0\ & \sigma^2_{\clu_x\clu_x} & 0 & 0 \\
            0 & \sigma^2_{\kin_x\clu_x} & 0\ & 0 & \sigma^2_{\clu_x\clu_x} & 0 \\
            0 & 0 & \sigma^2_{\kin_z\clu_z} & 0 & 0 & \sigma^2_{\clu_z\clu_z}
        \end{array}\right).
\eea
Notice now that each $3\times3$ block matrix is no longer proportional to the identity matrix,
so cannot be written as a Kronecker product.
In particular, we now see that the dispersions of the $x$- and
$y$-components differ from that of the $z$-component. Because of this
$\sigma^2_{\clu_x\clu_x}=\sigma^2_{\clu_y\clu_y}\neq\sigma^2_{\clu_z\clu_z}$ and
$\sigma^2_{\kin_x\kin_x}=\sigma^2_{\kin_y\kin_y}\neq\sigma^2_{\kin_z\kin_z}$; similarly,
for the cross-terms
$\sigma^2_{\kin_x\clu_x}=\sigma^2_{\kin_y\clu_y}\neq\sigma^2_{\kin_z\clu_z}$.
The probability distribution $p(\Dtot)$ is now given in terms of six independent quantities,
$\sigma^2_{\kin_x\kin_x}$, $\sigma^2_{\kin_z\kin_z}$,
$\sigma^2_{\clu_x\clu_x}$, $\sigma^2_{\clu_z\clu_z}$, $\sigma^2_{\kin_x\clu_x}$, and $\sigma^2_{\kin_z\clu_z}$.

We still have azimuthal
symmetry (rotations about the $z$-axis), and this is reflected in the $x$
and $y$ components. This is also seen in the multiplicity of the eigenvalues,
which are now $\lambdaup_1=1-2\sp t\Delta^2_z$ (multiplicity one),
$\lambdaup_2=1-2\sp t\Delta^2_x$ (multiplicity two), and $\lambdaup_3=1$ (multiplicity three),
where $\Delta_x^2$ and $\Delta_z^2$ are given by equations~\eqref{eqn:Delta_x} and \eqref{eqn:Delta_z},
respectively. With respect to the Galactic plane, we decompose into transverse ($x$ and $y$)
and longitudinal components ($z$), labelled collectively by
`$\perp$' and `$\|$', respectively.

We now have two distinct $t$-dependent eigenvalues (where previously in the full-sky
case there was only one), thus producing one additional pole in the
moment generating function:
\be
    {M}_{Y}(t)={M}_{Y_\perp}(t)\ssp{M}_{Y_\|}(t),
\ee
with ${M}_{Y_\perp}(t)\equiv{M}_{Y_x}(t)\ssp{M}_{Y_y}(t)$,
${M}_{Y_x}(t)={M}_{Y_y}(t)=(1-2\sp t\Delta^2_x)^{-1/2}$,
and ${M}_{Y_\|}(t)={M}_{Y_z}(t)=(1-2\sp t\Delta^2_z)^{-1/2}$ (these can be found by considering the
eigenvalues of the covariance matrix~\eqref{eqn:cov-6x6-cutsky}).
The PDFs of each of these constituent moment generating functions are obtained by
the following inverse Laplace transforms:
\begin{subequations}
\bea
    p_\perp(Y)
        &=\mathscr{L}^{-1}\big[{M}_{Y_\perp}(-t)\big](Y)=\frac{\rme^{-Y/(2\Delta^2_x)}}{2\sp\Delta^2_x}, \\
    p\Z{\|}(Y)
        &=\mathscr{L}^{-1}\big[{M}_{Y_\|}(-t)\big](Y)=\frac{\rme^{-Y/(2\Delta^2_z)}}{\big(2\sp\pi\sp Y\Delta^2_z\big)^{1/2}}.
\eea
\end{subequations}
By the convolution theorem of the unilateral Laplace transform, the PDF of the product
of moment generating functions $p(Y)=\mathscr{L}^{-1}[M_{Y_\perp}(-t)\ssp M_{Y_\|}(-t)]$ is
equal to the convolution $\mathscr{L}^{-1}[M_{Y_\perp}(-t)]\sp*\sp\mathscr{L}^{-1}[M_{Y_\|}(-t)]$,
which evaluates to
\bea
    p(Y)
        &=\int^Y_0\dif Y' p_\perp(Y-Y')\sp p\Z{\|}(Y') 
        =\frac{1}{\sqrt{\sp\varepsilon}}\, {\rm erf}
            \bigg(\frac{\varepsilon Y}{2\Delta^2_z}\bigg)
            \frac{\rme^{-Y/(2\Delta^2_x)}}{2\Delta^2_x}, \nonumber
\eea
where $\varepsilon\equiv1-\Delta^2_z/\Delta^2_x$ gives the `eccentricity' of the ellipsoid spanned by the eigenvectors. The use of the error function in
the foregoing equation is valid for $\varepsilon>0$; for $\varepsilon<0$, which is the
case for the Galactic plane cut, we have instead
\bea
    p(Y)
        &=\frac{2}{\sqrt{\pi|\varepsilon|}}\ssp \rme^{|\varepsilon|Y/(2\Delta^2_z)}\sp
            F\bigg(\frac{\sqrt{|\varepsilon|}\sqrt{Y}}{\sqrt2\Delta_z}\bigg)\sp
            \frac{\rme^{-Y/(2\Delta^2_x)}}{2\Delta^2_x},
\eea
where $F(x)\equiv\rme^{-x^2}\int^x_0\dif t\,\rme^{t^2}$ is the Dawson function~\citep[][\href{https://dlmf.nist.gov/7.2.5}{7.2.5}]{DLMF}.

The probability distribution of the amplitude $\Dtot=\sqrt{\sp Y}$,
in the presence of a Galactic plane cut, is now obtained easily from ${p(\Dtot)=p(Y\!=\!\Dtot^2)\ssp|\sp\dif Y/\dif\Dtot|}$;
it is
\be\label{eqn:pD-cutsky-pre}
    p(\Dtot)\ssp\dif\Dtot
        =\frac{2\Dtot\ssp\dif\Dtot}{\sqrt{\pi|\varepsilon|}\sp\Delta^2_x}\ssp
            F\bigg(\frac{\sqrt{|\varepsilon|}\Dtot}{\sqrt2\Delta_z}\bigg)
            \exp\bigg(-\frac{\Dtot^2}{2\Delta^2_z}\bigg),
\ee
where we have used that $|\varepsilon|=\Delta^2_z/\Delta^2_x-1>0$, since for us $\varepsilon<0$.
We can bring equation~\eqref{eqn:pD-cutsky-pre} into a form
more closely resembling the full-sky case
(i.e.~the Maxwell--Boltzmann distribution)
by using $\Delta_x^2=\Delta_z^2/(1-\varepsilon)$ to rewrite it
in terms of $\Delta_z$ and $\varepsilon$.
After some algebra, and after defining the function
$\mathcal{F}(x)\equiv F(x)/x$, we arrive at our final form
[cf.~equation~\eqref{eqn:p-X}]:
\be\label{eqn:pD-cutsky}
\begin{split}
    p(\Dtot)\ssp\dif\Dtot
    &=(1-\varepsilon)\ssp     \mathcal{F}\bigg(\frac{\sqrt{|\varepsilon|}\Dtot}{\sqrt2\sp\Delta_z}\bigg)
     \frac{4}{\sqrt\pi}\sp\exp\bigg(\!-\frac{\Dtot^2}{2\Delta_z^2}\bigg)
     \frac{\Dtot\Zu2}{2\sp\Delta_z^2}\ssp\frac{\dif\Dtot}{\sqrt2\sp\Delta_z}.
\end{split}
\ee
This is equation~\eqref{eqn:p-Dtot-cutsky} in the main text.
As a consistency check, it is easy to verify using $\mathcal{F}(a\sqrt{\sp\varepsilon}\ssp)\to1$, as $\varepsilon\to0$, that the full-sky expression~\eqref{eqn:p-X}
derived in the previous section is recovered in the limit $\varepsilon\to0$.
This is the special case of statistical isotropy, for which $\Delta_x=\Delta_y=\Delta_z$.

The first three moments of the PDF~\eqref{eqn:pD-cutsky} are
\begin{subequations}
\bea
    &\langle\Dtot\rangle
        =\sqrt\frac{2}{\pi}
            \bigg[\Delta_z + \Delta_x\frac{{\rm arcsinh}\big(\sqrt{|\varepsilon|}\sp\big)}{\sqrt{|\varepsilon|}}\bigg], \\[2pt]
    &\langle\Dtot^2\rangle
        =\Delta^2_z + 2\Delta^2_x, \\[3pt]
    &\langle\Dtot^3\rangle
        =\sqrt\frac{2}{\pi}\bigg[2\Delta^3_z+3\Delta_z\Delta^2_x
            +3\Delta^3_x\frac{{\rm arcsinh}\big(\sqrt{|\varepsilon|}\sp\big)}{\sqrt{|\varepsilon|}}\bigg].
\eea
\end{subequations}
In the limit $\varepsilon\to0$, these recover the usual moments of the Maxwell--Boltzmann distribution, i.e.\
$\langle\Dtot\rangle=\sqrt{\sp8/\pi}\sp\Delta$, $\langle\Dtot^2\rangle=3\Delta^2$, and
$\langle\Dtot^3\rangle=4\sqrt{\sp8/\pi}\sp\Delta^3$, with $\Delta=\Delta_x=\Delta_y=\Delta_z$.

\section{Dipole statistics on the cut sky}\label{app:galactic-cut}
In order to compare the dipole estimate with the theoretical value we will
convolve the full-sky predictions with the mask.

\subsection{Galactic plane cut\label{app:gal-cut}}
We make a {symmetric} cut about the Galactic plane, excluding a central region $|b|\leq b_{\rm cut}$.
Here $b_{\rm cut}=\ang{30}$, as in S21. This conservative cut removes half of the sky ($f_\mrm{sky}=0.5$). Since the
additional masks (accounting for poor-quality photometry, image artefacts, etc) covers merely $1.2\%$ of
the sky we will ignore their effect on the power, which principally affects the small angular
scales.
We will thus apply to the number-count field $\dif N/\dif\Omega(\n)$
the following mask:
\be
    \mathcal{W}(\vartheta)
        =
        \begin{cases}
            0, & \vartheta_-\leq\vartheta\leq\vartheta_+, \\
            1, & \text{otherwise},
        \end{cases}
\ee
where $\vartheta_\pm=\pi/2\pm b_{\rm cut}$. 
In terms of the Heaviside step function $\Theta$, this can also be written as
\be
    \mathcal{W}(\n)
        =\Theta(\n\cdot\hat{\mathbf{z}}-\mu\Z-)
        +\Theta(\mu\Z+-\n\cdot\hat{\mathbf{z}}),
\ee
with $\mu_{\pm}=\cos\vartheta_{\pm}$, and where
the first and second term selects the northern and southern caps, respectively.
Because $\mathcal{W}(\n)$ possesses azimuthal symmetry, its spherical harmonic expansion
is given purely in terms of the zonal modes ($m=0$):
$\mathcal{W}(\vartheta)=\sum_{\ell} \mathcal{W}_{\ell 0}\ssp Y_{\ell0}(\vartheta)$.
Furthermore, because $\mathcal{W}(\n)$ is parity symmetric,
$\mathcal{W}_{\ell m}=\int\dif^2\n~\mathcal{W}(\n)\sp Y^*_{\ell m}(\n)$ is equal to zero
for odd $\ell$;  for even $\ell$ these coefficients are non-zero and are
given by~\citep{Dahlen:2007sv}
\bea
    \mathcal{W}_{\ell m}
        &=2\sp\delta\Zu{\mrm{K}}_{m0}\ssp\sqrt{\frac{\pi}{2\ell+1}}\ssp
            \big[\mathcal{L}\Z{\ell-1}(\mu\Z-)-\mathcal{L}\Z{\ell+1}(\mu\Z-)\big]
         \quad\text{(even $\ell$)},
\eea
where $\calL_\ell(\mu)$ is the Legendre polynomial of degree $\ell$ [note $\mathcal{L}\Z{-1}(\mu)\equiv1$].
With $b_\mrm{cut}=\ang{30}$ we have for the first five non-zero coefficients
$\mathcal{W}_{00}\approx 1.77245$,
$\mathcal{W}_{20}\approx 1.48625$,
$\mathcal{W}_{40}\approx-0.623128$,
$\mathcal{W}_{60}\approx-0.131059$, and
$\mathcal{W}_{80}\approx 0.422182$. 

\subsection{Mode coupling}\label{app:mode-coupling}
A cut on the sky breaks statistical
isotropy. In harmonic space this implies that the covariance
matrix of $a\Z{\ell m}$'s can have off-diagonal components, i.e.~we no longer have that
$\langle{a}\Z{\ell m}\sp {b}_{\ell'm'}^* \rangle=C_\ell\ssp\delta^\mrm{K}_{\ell\ell'}\sp\delta^\mrm{K}_{mm'}$.
Now, $a_{\ell m}$ with different $\ell$
or $m$ may be correlated.

We now apply the mask $\mathcal{W}(\n)$ to the number-count
field
$\dif N/\dif\Omega(\n)\to\mathcal{W}(\n)\dif N/\dif\Omega(\n)$.
Denoting by $\tilde{a}_{\ell m}$
and $\tilde{b}_{\ell m}$ the coefficients of the masked fields
$\mathcal{W}(\n)\sp\dqso(\n)$ and $\mathcal{W}(\n)(\dkin\cdot\n)$,
respectively, it can be shown using the properties of spherical
harmonics that we now have a linear coupling of the
underlying $a_{\ell m}$ and $b_{\ell m}$~\citep{1980lssu.book.....P}:
\be
    \tilde{a}\Z{\ell m}
        =\sum_{\ell' m'} \mathcal{A}\Z{\ell m\ell'm'}\,\sp a\Z{\ell'm'},\qquad
    \tilde{b}\Z{\ell m}
        =\sum_{\ell' m'} \mathcal{B}\Z{\ell m\ell'm'}\,\sp b\Z{\ell'm'}.
\ee
Here $\mathcal{A}_{\ell m\ell' m'}$ is a harmonic kernel which is
determined by the window function $\mathcal{W}(\n)$; for a general
$\mathcal{W}(\n)=\sum_{\ell,m} \mathcal{W}_{\ell m}Y_{\ell m}(\n)$, it
is given by~\citep{Hivon:2002jp}
\bea
    \mathcal{A}_{\ell_1 m_1 \ell_2 m_2}
        &\equiv\sum_{\ell_3,m_3}\mathcal{W}_{\ell_3 m_3}\int\dif^2\n\ Y_{\ell_1 m_1}(\n)\, Y\Zu{*}_{\ell_2m_2}(\n)\, Y_{\ell_3m_3}(\n) \nonumber\\
        &=(-1)^{m_2}\sum_{\ell_3,m_3}\sp \mathcal{W}\Z{\ell_3 m_3}
            \bigg[\frac{(2\ell_1+1)(2\ell_2+1)(2\ell_3+1)}{4\pi}\bigg]^{1/2}\! \nonumber\\[2pt]
        &\qquad\qquad\quad
            \times
            \begin{pmatrix}
                \ell_1 & \ell_2 & \ell_3 \\
                0 & 0 & 0
            \end{pmatrix}
            \begin{pmatrix}
                \ell_1 & \ell_2 & \ell_3 \\
                m_1 & -m_2 & m_3
            \end{pmatrix},
       \label{eqn:A-wigner}
\eea
where the arrays are Wigner $3j$ symbols,
coefficients that encode selection rules for the allowed combinations of $\ell$'s and $m$'s.
Using that $\mathcal{A}_{\ell m\ell'm'}=\mathcal{A}_{\ell m\ell'm}\ssp\delta^\mrm{K}_{mm'}$ (by azimuthal symmetry), and
$\mathcal{A}_{\ell m\ell'm'}^*=\mathcal{A}_{\ell m\ell'm'}$ (because $\mathcal{W}^*_{\ell0}=\mathcal{W}_{\ell0}$),
we have
for two arbitrary fields, each with azimuthally-symmetric masks (which need not be identical),
\bea
    \big\langle\tilde{a}^{\phantom{*}}\Z{\ell_1m_1}\tilde{b}\Z{\ell_2m_2}^*\big\rangle
        &=\sum_{\ell_1',\ell_2',m_1',m_2'} \mathcal{A}_{\ell_1m_1\ell_1'm_1'}\ssp \mathcal{B}_{\ell_2m_2\ell_2'm_2'}^*\ssp
            \big\langle a_{\ell_1'm_1'}\sp b^*_{\ell_2'm_2'}\big\rangle \nonumber\\
        &=\delta^\mrm{K}_{m_1m_2}\sum_{\ell} \mathcal{A}_{\ell_1m_1\ell m_1}\ssp
            \mathcal{B}_{\ell_2m_2\ell m_1}\sp C_{\ell}.
\eea

In our case -- with $\tilde{a}_{\ell m}$ corresponding to the clustering field and $\tilde{b}_{\ell m}$
corresponding to the kinematic field -- we only need to consider $\ell_1=\ell_2=1$, with identical mask 
applied to both fields ($\mathcal{A}=\mathcal{B}$).
In terms of the underlying power spectra we have
\begin{subequations}\label{eqn:power-masked}
\bea
    \big\langle\tilde{a}^{\phantom{*}}\Z{1m}\tilde{b}\Z{1m'}^*\big\rangle
        &=\delta^\mrm{K}_{mm'}\sp|\mathcal{A}_{1m1m}|^2\sp C\Zu{\ssp\kin\clu}_1, \\[5pt]
    \big\langle\tilde{a}^{\phantom{*}}\Z{1m}\tilde{a}\Z{1m'}^*\big\rangle
        &=\delta^\mrm{K}_{mm'}\sum_{\ell\ \text{odd}} |\mathcal{A}_{1m\ell m}|^2 C\Zu{\ssp\clu\clu}_{\ell}, \\
    \big\langle\tilde{b}^{\phantom{*}}\Z{1m}\tilde{b}\Z{1m'}^*\big\rangle
        &=\delta^\mrm{K}_{mm'}\sp|\mathcal{A}_{1m1m}|^2\sp C\Zu{\ssp\kin\kin}_1,
\eea
\end{subequations}
where only odd $\ell$ need to be considered because of the selection rules of Wigner $3j$ symbols in equation~\eqref{eqn:A-wigner}.
Note that no off-diagonal entries are induced in the
covariance matrix in harmonic space. However, in Cartesian space the covariance matrix
is no longer proportional to the identity matrix (although it remains diagonal).
Though we will no longer have spherical symmetry we still have azimuthal symmetry.

For $b_\mrm{cut}=\ang{30}$, the first three coupling coefficients are
$|\mathcal{A}_{1010}|^2\approx 0.76563$,
$|\mathcal{A}_{1030}|^2\approx 0.046143$, and
$|\mathcal{A}_{1050}|^2\approx 0.034272$.
Also
$|\mathcal{A}_{1111}|^2\approx 0.0976563$,
$|\mathcal{A}_{1131}|^2\approx 0.155731$, and
$|\mathcal{A}_{1151}|^2\approx 0.00955939$.
These values indicate a substantial loss of power for the $x$- and $y$-components of the dipole.

\subsection{Dipole dispersions}\label{app:stats-cut}
We now need to relate the statistics in harmonic space back to the statistics
in Cartesian space. To facilitate this it will be
convenient to define the complex-valued vectors
$\bm{\tilde{a}}\Z1\equiv(\tilde{a}_{1-1},\tilde{a}_{11},\tilde{a}_{10})\Zu\T$
and $\bm{\tilde{b}}\Z1\equiv(\tilde{b}_{1-1},\tilde{b}_{11},\tilde{b}_{10})\Zu\T$.
We then have the following linear relation between the Cartesian- and harmonic-space dipoles:
\be
    \dint=\sqrt{\frac{3}{4\pi}}\,\mat{U}^\dagger\ssp\bm{\tilde{a}}\Z1, \quad
    \dkin=\sqrt{\frac{3}{4\pi}}\,\mat{U}^\dagger\ssp\bm{\tilde{b}}\Z1,
\ee
where the tilde indicates coefficients of masked fields, and $\mat{U}$ is given by equation~\eqref{eq:U}.
With these relations we can express the covariances of these vectors,
$\langle\mbf{d}_\mrm{X}^{\phantom\T}\mbf{d}_\mrm{Y}^\T\rangle$, in terms of the
underlying power spectra:
\begin{subequations}
\bea
    \big\langle\dkin\Zu{\phantom\T}\dkin\Zu\T\big\rangle
        &=\frac{3}{4\pi}\,\mathbf{U}^\dagger\langle\bm{\tilde{b}}_1^{\phantom\dagger}\bm{\tilde{b}}_1^\dagger\rangle\mathbf{U} \nonumber\\[-6pt]
        &={\rm diag}\big(\sp|\mathcal{A}_{1111}|^2,\sp\,|\mathcal{A}_{1111}|^2,\sp\,|\mathcal{A}_{1010}|^2\big)\sp\frac{3C_1\Zu{\ssp\kin\kin}}{4\pi}, \\
    \big\langle\dkin\Zu{\phantom\T}\dint\Zu\T\big\rangle
        &=\frac{3}{4\pi}\,\mathbf{U}^\dagger\langle\bm{\tilde{b}}_1^{\phantom\dagger}\bm{\tilde{a}}_1^\dagger\rangle\mathbf{U} \nonumber\\[-6pt]
        &={\rm diag}\big(\sp|\mathcal{A}_{1111}|^2,\sp\,|\mathcal{A}_{1111}|^2,\sp\,|\mathcal{A}_{1010}|^2\big)\sp\frac{3C_1\Zu{\ssp\kin\clu}}{4\pi}, \\
    \big\langle\dint\Zu{\phantom\T}\dint\Zu\T\big\rangle
        &=\frac{3}{4\pi}\,\mat{U}^\dagger\langle\bm{\tilde{a}}_1^{\phantom\dagger}\bm{\tilde{a}}_1^\dagger\rangle\mat{U} \nonumber\\[-4pt]
        &=\sum_{\ell\ \text{odd}}
         {\rm diag}\big(\sp|\mathcal{A}_{11\ell1}|^2,\sp\,|\mathcal{A}_{11\ell1}|^2,\sp\,|\mathcal{A}_{10\ell0}|^2\big)\sp\frac{3C_\ell\Zu{\ssp\clu\clu}}{4\pi},
\eea
\end{subequations}
The block-matrix form of the $6\times6$ cut-sky covariance~\eqref{eqn:cov-6x6-cutsky} is constructed
out of these $3\times3$ covariances.
The first matrix $\langle\dkin\Zu{\phantom\T}\dkin\Zu\T\rangle\propto\mathbf{I}_3$ is unchanged from
before; the second and third matrices are diagonal but have different dispersions from before. For instance,
the dispersion of the $z$-component is proportional to
$\langle|\tilde{a}\Z{10}|^2\rangle=\sum_\ell|\mathcal{A}_{10\ell0}|^2\sp C^{}_\ell$, while the $x$- and
$y$-components both have dispersions proportional to $\langle|\tilde{a}\Z{11}|^2\rangle=\sum_\ell|\mathcal{A}_{11\ell1}|^2\sp C^{}_\ell$,
with $\mathcal{A}_{11\ell1}\neq \mathcal{A}_{10\ell0}$.

We can relate the harmonic form to the dispersions in the cut-sky covariance matrix~\eqref{eqn:cov-6x6-cutsky}
as follows. Again, because of azimuthal symmetry, we recall that the dispersions for the $x$- and
$y$-components are equal (though modified from before),
e.g.~$\sigma^2_{\clu_x\clu_x}=\sigma^2_{\clu_y\clu_y}$, $\sigma^2_{\kin_x\clu_x}=\sigma^2_{\kin_y\clu_y}$, etc.
The new dispersions, in terms of the spectra, now read
\bea
    \sigma\Zu2\Z{\kin_x\kin_x}
        &=\frac{3}{4\pi}\ssp|\mathcal{A}_{1111}|^2\ssp C\Zu{\ssp\kin\kin}_1,
    &&\sigma\Zu2\Z{\kin_z\kin_z}
        =\frac{3}{4\pi}\ssp|\mathcal{A}_{1010}|^2\ssp C\Zu{\ssp\kin\kin}_1, \nonumber\\
    \sigma\Zu2\Z{\clu_x\clu_x}
        &=\frac{3}{4\pi}\sum_{\ell\ \text{odd}}|\mathcal{A}_{11\ell1}|^2\ssp C\Zu{\ssp\clu\clu}_\ell,
    &&\sigma\Zu2\Z{\clu_z\clu_z}
        =\frac{3}{4\pi}\sum_{\ell\ \text{odd}}|\mathcal{A}_{10\ell0}|^2\ssp C\Zu{\ssp\clu\clu}_\ell, \nonumber\\
    \sigma\Zu2\Z{\kin_x\clu_x}
        &=\frac{3}{4\pi}\ssp|\mathcal{A}_{1111}|^2\ssp C\Zu{\ssp\kin\clu}_1,
    &&\sigma\Zu2\Z{\kin_z\clu_z}
        =\frac{3}{4\pi}\ssp|\mathcal{A}_{1010}|^2\ssp C\Zu{\ssp\kin\clu}_1, \nonumber
\eea
where $C\Zu{\mrm{XY}}_\ell$ is given by equation~\eqref{eqn:Cab-ell}.
The dispersions related to the total dipole are given in terms of these by
equations~\eqref{eqn:Delta_x} and \eqref{eqn:Delta_z}.
For \LCDM, though we have $C\Zu{\ssp\clu\clu}_\ell>C\Zu{\ssp\clu\clu}_1$, for $\ell\geq2$, the
leakage of power into the dipole is not enough to overcome the loss of power
across all $\ell$, e.g.~$\sigma^2_{\kin_z\kin_z}=|A_{1010}|^2\sigkin^2\approx0.77\sigkin^2$, a
$23\%$ loss of kinematic dipole power.
Because of this we find that we only need to sum up the first few odd multipoles to attain
convergence (here we take $\ell_\mrm{max}=9$).

\bsp	
\label{lastpage}
\end{document}